\documentclass[12pt]{article}
\usepackage{epsfig}
\usepackage{amssymb}
\newcommand{\DE}{\relax{\rm I\kern-.18em E}}
\newcommand{\IP}{\relax{\rm I\kern-.18em P}}
\newcommand{\IR}{\relax{\rm I\kern-.18em R}}
\newcommand{\IB}{\relax{\rm I\kern-.18em B}}
\newcommand{\be}{\begin{equation}}
\newcommand{\ee}{\end{equation}}
\newcommand{\ben}{\begin{eqnarray}\displaystyle}
\newcommand{\een}{\end{eqnarray}}
\newcommand{\sectiono}[1]{\section{#1}\setcounter{equation}{0}}

\newdimen\tableauside\tableauside=1.0ex
\newdimen\tableaurule\tableaurule=0.4pt
\newdimen\tableaustep
\def\phantomhrule#1{\hbox{\vbox to0pt{\hrule height\tableaurule width#1\vss}}}
\def\phantomvrule#1{\vbox{\hbox to0pt{\vrule width\tableaurule height#1\hss}}}
\def\sqr{\vbox{%
  \phantomhrule\tableaustep
  \hbox{\phantomvrule\tableaustep\kern\tableaustep\phantomvrule\tableaustep}%
  \hbox{\vbox{\phantomhrule\tableauside}\kern-\tableaurule}}}
\def\squares#1{\hbox{\count0=#1\noindent\loop\sqr
  \advance\count0 by-1 \ifnum\count0>0\repeat}}
\def\tableau#1{\vcenter{\offinterlineskip
  \tableaustep=\tableauside\advance\tableaustep by-\tableaurule
  \kern\normallineskip\hbox
    {\kern\normallineskip\vbox
      {\gettableau#1 0 }%
     \kern\normallineskip\kern\tableaurule}%
  \kern\normallineskip\kern\tableaurule}}
\def\gettableau#1{\ifnum#1=0\let\next=\null\else
\squares{#1}\let\next=\gettableau\fi\next}

\newcommand{\figref}[1]{Fig.~\protect\ref{#1}}

\begin{document}
{}~
\hfill\vbox{\hbox{hep-th/0210145}
\hbox{HUTP-02/A052}
}\break

\vskip .6cm

\centerline{\Large \bf
Enumerative geometry and knot invariants}

\medskip

\vspace*{4.0ex}

\centerline{\large \rm
Marcos Mari\~no }

\vspace*{4.0ex}

\centerline{  Jefferson Physical Laboratory, Harvard University}
\centerline{  Cambridge MA 02138, USA}
\vspace*{2.0ex}
\centerline{marcos@born.harvard.edu}

\vspace*{5.0ex}

\centerline{\bf Abstract} \bigskip

We review the string/gauge theory duality relating
Chern-Simons theory and topological strings on noncompact Calabi-Yau
manifolds, as well as its mathematical implications for
knot invariants and enumerative geometry.

\vfill

\eject

\baselineskip=16pt
\pagestyle{plain}
\tableofcontents

\sectiono{Introduction}
Enumerative geometry and knot theory have benefitted considerably from the
insights and results in string theory and topological field theory. The
theory of Gromov-Witten invariants has emerged mostly from the
consideration of topological sigma models and topological strings, and
mirror symmetry has provided a surprising point of view with powerful
techniques and deep implications for the theory of enumerative
invariants. On the other hand, the new invariants of knots and links
that emerged in the eighties turned out to be deeply related to
Chern-Simons
theory, a
topological gauge theory introduced by Witten in \cite{cs}, which also
provided a new family of invariants of three-manifolds. It is safe to say
that these two topics, enumerative geometry and knot theory, have been
deeply transformed through the emergence of these connections to physics.

A more recent surprise, however, is that, in
many situations, knot invariants are related to enumerative
invariants. The reason is that Chern-Simons gauge theory has a string
description in the sense envisaged by 't Hooft \cite{thooft},
and this description turns
out to involve topological strings, {\it i.e.} the physical counterparts of
Gromov-Witten invariants. This relation between two seemingly
unrelated areas of geometry is therefore based
on a beautiful realization of the large $N$
string/gauge theory duality. The connection between Chern-Simons theory and
topological strings was first pointed out by Witten in \cite{csts}, and the
current picture emerged in the works of Gopakumar and Vafa \cite{gvgeom}
and Ooguri and Vafa \cite{ov}.

In this paper we have tried to review these developments. We have
focused mostly in presenting results, general ideas and examples.
Some of the physical arguments leading to these results
are not covered in detail, mostly for reasons of space, but also
with the hope that mathematicians will find this review more
readable. Important related developments, like the interplay with
mirror symmetry and the relation with M-theory on manifolds of
$G_2$ holonomy, are only mentioned in the text. Other reviews of
the topics covered here include \cite{vafarev,labasrev}, and more
recently \cite{grassi}, which provides extensive mathematical
background.

The plan of this paper is the following. In section 2 we review some
basic facts about open and closed topological strings and their structure
in terms of integer invariants. In section 3, we give a quick review of
Chern-Simons theory and knot and link invariants. In section 4, we state the
basic ideas of string/gauge theory duality in the $1/N$ expansion,
and we show, following
Gopakumar and Vafa,
that Chern-Simons theory has a description in terms of closed strings
on the resolved conifold. In section
5 we show in detail how to incorporate Wilson loops in the duality. It turns out
that the Chern-Simons/string duality can be extended to closed
strings propagating in more complicated toric geometries, and we summarize some
of the results in section 6. Finally, some conclusions and open problems are
collected in section 7.

\sectiono{Topological strings}

\subsection{Topological sigma models}

The starting point to construct topological strings is
an ${\cal N}=(2,2)$ superconformal field theory, the
${\cal N}=(2,2)$ nonlinear sigma model. This model can be twisted
in two ways in order to produce a topological field theory
\cite{topmodel,topmatter,tftmirror}, which are usually called the
A and the B model. We will focus here on the A-model.

The field content of this model is the following. First, we
have a map $x: \Sigma_g \rightarrow X$ from a Riemann surface of genus
$g$ to a target space $X$, that will be a K\"ahler manifold of complex
dimension $d$. We also have fermions $\chi \in x^*(TX)$, which are
scalars on $\Sigma_g$, and a fermionic one form $\psi_{\alpha}$ with
values in $x^*(TX)$. This last field satisfies a selfduality condition
which implies that its only nonzero components are $\psi_{\bar z}^I
\in x^*(T^{(1,0)}X)$ and $\psi_{ z}^{\overline I}
\in x^*(T^{(0,1)}X)$, where $T^{(1,0)}X, T^{(0,1)}X$ denote, respectively,
the holomorphic and the antiholomorphic tangent bundles, and $I, \overline I$ are
the corresponding indices. The theory also has a BRST, or topological, charge
$Q$ which acts on the fields according to
\begin{eqnarray}
\{ Q, x \} &=& i\chi, \nonumber\\
\{ Q, \chi \} &=&0, \nonumber\\
\{ Q, \psi_{\bar z}^I \} &=&-\partial_{\bar z}x^I -i \chi^J
\Gamma_{JK}^I \psi_{\bar z}^K, \nonumber\\
\{ Q, \psi_z^{\overline I} \} &=&-\partial_z x^{\overline I} -i
\chi^{\overline J}
\Gamma_{{\overline J} {\overline K}}^{\overline I}
\psi_{\bar z}^{\overline K}. \nonumber\\
\label{qtrans}
\end{eqnarray}
The twisted Lagrangian turns out to be $Q$-exact, up to a topological term:
\begin{equation}
{\cal L}= i \{ Q, V\} + \int_{\Sigma_g} x^*(\omega),
\label{toplag}
\end{equation}
where $\omega=J + iB$ is the complexified K\"ahler class of $X$, and $V$
(sometimes called the gauge fermion) is given by
\begin{equation}
V=\int_{\Sigma_g} d^2z\, G_{I {\overline J}} ( \psi_z^{\overline I}
\partial_{\bar z} x^J + \partial_x x^{\overline I} \psi_{\bar z}^J).
\label{gaugef}
\end{equation}
In this equation, $G_{I {\overline J}}$ is the K\"ahler metric
of $X$. Notice that the last term in (\ref{toplag}) is a topological invariant
characterizing the homotopy type of the map $x: \Sigma_g \rightarrow X$,
therefore the energy-momentum tensor of this theory is given by:
\begin{equation}
T_{\alpha \beta}= \{Q, b_{\alpha \beta} \},
\label{qex}
\end{equation}
where $b_{\alpha \beta}= \delta V/\delta g^{\alpha \beta}$. The fact that
the energy-momentum tensor is $Q$-exact means that the theory is
topological, and the fact that the Lagrangian is $Q$-exact up to a
topological term means that the
semiclassical approximation is exact. The classical solutions of the sigma
model action are holomorphic maps $x:
\Sigma_g \rightarrow X$, which are also known as
worldsheet instantons, and the functional integral
localizes to these configurations. The relevant operators in this theory, as in any
topological theory of cohomological type, are the $Q$-cohomology
classes. In this case they are given by operators of the form,
\begin{equation}
{\cal O}_{\phi}=\phi_{i_1 \cdots i_p} \chi^{i_1} \cdots \chi^{i_p},
\label{qops}
\end{equation}
where $\phi=\phi_{i_1 \cdots i_p}dx^{i_1}\wedge \cdots \wedge
dx^{i_p}$ is a closed
$p$-form representing a nontrivial class in $H^p(X)$. Moreover, one can
derive a selection rule for correlation functions of such operators:
the vacuum expectation value $\langle {\cal O}_{\phi_1} \cdots
{\cal O}_{\phi_\ell} \rangle$ vanishes unless
\begin{equation}
\sum_{k=1}^{\ell} {\rm deg}({\cal O}_{\phi_k})=
2d(1-g) + 2\int_{\Sigma_g} x^*(c_1(X)),
\label{saturate}
\end{equation}
where ${\rm deg}({\cal O}_{\phi_k})={\rm deg}(\phi)$.
The right hand of this equation is nothing but the virtual dimension of
the moduli space of holomorphic maps, ${\cal M}^{\rm hol}_{\Sigma_g
\rightarrow X}$. Since the operators (\ref{qops}) can be interpreted as
differential forms on this moduli space, the above selection rule just says
that we have to integrate top forms.

In the case of a Calabi-Yau manifold of complex dimension 3, we
have $c_1(X)=0$, and the selection rule says that at genus $g=0$
({\it i.e.} when the Riemann surface is a sphere ${\bf S}^2$) we
have to insert three operators associated to 2-forms. The
correlation functions can be evaluated by summing over the
different topological sectors of holomorphic maps. These sectors
can be labelled by ``instanton numbers.'' Let $\Sigma_i$ denote a
basis of $H_2(X)$, with $i=1, \cdots, b_2$. If the image of
$x({\bf S}^2)$ is in the homology class $\beta=\sum_i n_i
\Sigma_i$, then we will say that the worldsheet instanton is in
the sector specified by $\beta$, or equivalently, by the integers
$n_i$. The trivial sector corresponds to $\beta=0$, {\it i.e.} the
image of the sphere is a point in the target, and in this case the
correlation function is just the classical intersection number
$D_1 \cap D_2 \cap D_3$ of the three divisors $D_i$, $i=1,2,3$,
associated to the 2-forms, while the nontrivial instanton sectors
give an infinite series. The final answer looks, schematically,
\begin{equation}
\langle {\cal O}_{\phi_1}{\cal O}_{\phi_2}
{\cal O}_{\phi_3} \rangle=(D_1 \cap D_2 \cap D_3) + \sum_{\beta}
I_{0,3, \beta}(\phi_1,\phi_2, \phi_3)
q^{\beta}
\label{threepoint}
\end{equation}
The notation is as follows: let $\omega = \sum_{i=1}^{b_2} t_i \omega_i$,
be the complexified K\"ahler form of $X$,
where $\omega_i$ is a basis for $H^2(X)$ dual to $\Sigma_i$,
and $t_i$ are the complexified
K\"ahler parameters. Set $q_i={\rm e}^{-t_i}$.
If $\beta=\sum_i n_i \Sigma_i$, then $q^{\beta}$ denotes $\prod_i
q_i^{n_i}$. The coefficients
$I_{0,3, \beta}(\phi_1,\phi_2, \phi_3)$ ``count'' in some appropriate way
the number of holomorphic maps from the sphere to the Calabi-Yau, in the
topological sector specified by $\beta$, and in such a
way that the point of insertion of ${\cal O}_{\phi_i}$ gets mapped to the
divisor $D_i$. This is an example of a Gromov-Witten invariant, although
to get the general picture we have to couple the model to gravity, as we
will
see very soon.

When $c_1(X)>0$, correlation functions also have the structure of
(\ref{threepoint}): the trivial sector gives just the classical
intersection number of the cohomology ring, and then there are
quantum corrections associated to the worldsheet instantons. One
important aspect of the case $c_1(X)>0$ is that the right hand
side of (\ref{saturate}) contains the positive integer $\sum_i n_i
\int_{\Sigma_i} c_1(X)$, where $n_i$ are the instanton numbers
labelling the topological sector of the holomorphic map. As the
$n_i$ increase, it won't be possible to satisfy the selection rule
for the insertions. Therefore, only a finite number of topological
sectors contribute to the correlation function, which will be
given by the sum of a classical intersection number plus a finite
number of ``quantum'' corrections. This is the starting point in the 
definition of the quantum cohomology of $X$, see \cite{ck} for details.

\subsection{Closed topological strings}

In the above considerations on topological sigma models we have focused on
 $g=0$. For $g=1$ and a
Calabi-Yau manifold, the only vacuum expectation value (vev) that
may lead to a nontrivial answer is
that of the unit operator, {\it i.e.} the partition function itself, while
for $g>1$ the virtual dimension of the moduli space is negative and the
above theory is no longer useful to study the enumerative geometry of the
target space $X$. This
corresponds mathematically to the fact that, for a generic metric on the
Riemann surface $\Sigma_g$, there are no holomorphic maps at genus
$g>1$. In order to circumvent this problem, we have to couple the theory to
two-dimensional gravity, which means considering all possible metrics on
the Riemann surface. The resulting model is called a {\it
topological string} theory. We will start by giving a general idea from a more
 mathematical point of view (see \cite{ck} for a rigorous
 discussion), and then we will
present the physical construction.

The moduli space of possible metrics (or equivalently, complex
structures) on a Riemann surface with punctures is the famous
Deligne-Mumford space ${\overline M}_{g,n}$ of stable curves with
$n$ marked points (the definition of what stable means can be
found for example in \cite{fm}). The moduli space we have to
consider in the theory of topological strings also involves maps.
It consists on one hand of a point in ${\overline M}_{g,n}$, {\it
i.e.} a Riemann surface with $n$ punctures, $(\Sigma_g, p_1,
\cdots, p_n)$, and this involves a choice of complex structure on
$\Sigma_g$. On the other hand, we have a map $x:\Sigma_g
\rightarrow X$ which is holomorphic with respect to the choice of
complex structure on $\Sigma_g$.

Let us now fix the topological sector of the holomorphic map, {\it i.e.} the homology
class $\beta=x_*[\Sigma_g]$. In general, there will be many maps in this
sector. The set
given by the possible data $(x,\Sigma_g, p_1, \cdots, p_n)$ associated to
the class $\beta$ can be promoted to a moduli space ${\overline
M}_{g,n}(X,\beta)$, provided a
certain number of conditions are satisfied. This is the basic moduli space
we will need in the theory of topological strings. Its (complex)
virtual dimension is
given by:
\begin{equation}
(1-g)(d -3) + n + \int_{\Sigma_g} x^*(c_1(X)).
\label{vmoduli}
\end{equation}
If we compare (\ref{vmoduli}) to (\ref{saturate}), we see that there is an
 extra $3(g-1) + n$ which comes from the Mumford-Deligne space
${\overline M}_{g,n}$.
 The moduli space ${\overline
M}_{g,n}(X,\beta)$ comes
equipped with the natural maps
\begin{eqnarray}
\pi_1:  {\overline
M}_{g,n}(X,\beta)& \longrightarrow & X^n,\nonumber\\
\pi_2:    {\overline
M}_{g,n}(X,\beta) &\longrightarrow &  {\overline M}_{g,n}.
\end{eqnarray}
The first map is easy to define: given a point $(x,\Sigma_g, p_1,
\cdots, p_n)$ in ${\overline M}_{g,n}(X,\beta)$, we just compute
$(x(p_1), \cdots, x(p_n))$. The second map sends $(x,\Sigma_g,
p_1, \cdots, p_n)$ to $(\Sigma_g, p_1, \cdots, p_n)$, {\it i.e.}
forgets the information about the map and leaves the punctured
curve (there are some subtleties with this map, associated to the
stability conditions; see \cite{ck}). We can now formally define
the Gromov-Witten invariant $I_{g,n,\beta}$ as follows. Let us
consider cohomology classes $\phi_1, \cdots, \phi_n$ in $H^*(X)$.
The map $\pi_1$ induces a map $\pi_1^*: H^*(X)^n \rightarrow
H^*({\overline M}_{g,n}(X,\beta))$, and we can pullback $\phi_1
\otimes \cdots \otimes \phi_n$ to get a differential form on the
moduli space of holomorphic maps. This form can be integrated as
long as there is a well-defined fundamental class for this space,
and the result is the Gromov-Witten invariant
$I_{g,n,\beta}(\phi_1, \cdots, \phi_n)$:
\begin{equation}
I_{g,n,\beta}(\phi_1, \cdots, \phi_n)=\int_{{\overline
M}_{g,n}(X,\beta)} \pi_1^* (\phi_1 \otimes \cdots \otimes \phi_n).
\label{gwinv}
\end{equation}
By using the Gysin map $\pi_{2!}$, one can reduce this to an integral over
the moduli space of curves ${\overline M}_{g,n}$. The Gromov-Witten
invariant $I_{g,n,\beta}(\phi_1, \cdots, \phi_n)$ vanishes unless the
degree of the form equals the dimension of the moduli space. Therefore, we
have the following selection rule:
\begin{equation}
{1 \over 2}\sum_{i=1}^n {\rm deg}(\phi_i)=(1-g)(d -3) + n +
\int_{\Sigma_g} x^*(c_1(X))
\label{degcons}
\end{equation}
Notice that Calabi-Yau threefolds play a special role in the theory,
since for those targets the virtual dimension only depends on the number of
punctures, and therefore the above condition is always satisfied if the
forms $\phi_i$ have degree 2. The invariants (\ref{gwinv})
generalize the invariants
obtained from topological sigma models. In particular, $I_{0,3, \beta}$ are
the invariants involved in the evaluation of correlation functions of the
topological sigma model with a Calabi-Yau threefold as its target in
(\ref{threepoint}). When $n=0$, one gets an
invariant $N_{g,\beta}=I_{g,0,\beta}$ which does not require any
insertions. We will refer to this as the Gromov-Witten invariant of the
Calabi-Yau threefold $X$ at genus $g$ and in the class $\beta$. These are the
only (closed) Gromov-Witten invariants that we will deal with here. It can
be also shown that, for genus $0$ \cite{ck},
\begin{equation}
I_{0,3, \beta}(\phi_1, \phi_2, \phi_3)=N_{0,\beta}\int_{\beta} \phi_1   \int_{\beta} \phi_2
\int_{\beta} \phi_3,
\label{geno}
\end{equation}
so from these Gromov-Witten invariants one can recover as well the
information about the three-point functions of the topological sigma model.

The physical point of view on the Gromov-Witten invariants $N_{g,\beta}$
comes about as follows. It is clear that we have to couple the topological
sigma model to two
dimensional gravity in order to get nontrivial invariants. To do that, one
realizes \cite{dvv,bcov} that the structure of the twisted theory is
tantalizingly close to that of the bosonic string. In the bosonic string, there is
a nilpotent BRST operator, $Q_{\rm BRST}$, and the energy-momentum tensor
turns out to be a $Q_{\rm BRST}$-commutator: $T(z)=\{Q_{\rm BRST}, b(z)
\}$. This is precisely the same structure that we found in (\ref{qex}), so
the field $b_{\alpha \beta}$ plays the role of a ghost. Therefore, one can
just follow the prescription of coupling to gravity for the bosonic string
and define a genus $g$ free energy as follows:
\begin{equation}
F_g= \int_{{\overline M}_{g}} \langle \prod_{k=1}^{6g-6} (b, \mu_k)
\rangle,
\label{fg}
\end{equation}
where
\begin{equation}
(b, \mu_k)=\int d^2 z (b_{zz}(\mu_k)_{\bar z}^{~z} + b_{\bar z \bar z}
({\overline \mu}_k)_{z}^{~\bar z}),
\end{equation}
and $\mu_k$ are the usual Beltrami differentials.
The vev in (\ref{fg}) refers to the path integral over the
fields of the twisted sigma model. The result, which depends on the choice
of complex structure of the Riemann surface, is then integrated over the
moduli space ${\overline M}_{g}$. $F_g$ can be evaluated again, like in the
topological sigma model, as a sum over instanton sectors. It turns out
\cite{bcov} that
$F_g$ is a generating functional for the Gromov-Witten invariants
$N_{g,\beta}$, or more precisely,
\begin{equation}
\label{fgt}
F_g(t)= \sum_{\beta}N_{g,\beta}q^{\beta}.
\end{equation}
It is also useful to introduce a generating functional for the all-genus
free energy:
\begin{equation}
F(g_s,t)=\sum_{g=0}^{\infty} F_g(t) g_s^{2g-2}.
\label{freen}
\end{equation}
The parameter $g_s$ can be regarded as a formal variable, but in the context
of type II strings it is nothing but the string coupling constant.

The first term in (\ref{fgt}) corresponds to the contribution of constant
maps, with $\beta=0$. It was shown in
\cite{bcov} (see also \cite{gp}) that, for $g\ge 2$,
this contribution can be
expressed as an integral over ${\overline M}_g$. The result is as follows:
on ${\overline M}_{g}$ there is a complex vector bundle $\DE$
of rank $g$, called the Hodge bundle, whose fiber at a point $\Sigma$ is
$H^0 (\Sigma, K_{\Sigma})$. The contribution of constant maps to $F_g$ is then
given by
\be
N_{g,0}=(-1)^g {\chi (X) \over 2} \int_{{\overline M}_g} c_{g-1}^3(\DE), \,\,\,\,\,\,\,\,
g \ge 2,
\label{hint}
\ee
where $c_{g-1}$ is the $(g-1)$-th Chern class of $\DE$, and $\chi(X)$ is
the Euler characteristic of the target space.

In general, Gromov-Witten invariants can be computed by using the
localization techniques pioneered by Kontsevich \cite{kont}. These 
techniques are easier to implement in the case of
non-compact Calabi-Yau manifolds (the so-called {\it local} case), where
one can compute $N_{g,\beta}$ for arbitrary genus. For example,
let us consider the non-compact Calabi-Yau manifold ${\cal O}(-3)
\rightarrow \IP^2$. This is the total space of $\IP^2$ together
with its anticanonical bundle, and it has $b_2=1$, corresponding
to the hyperplane class of $\IP^2$. Therefore, the class $\beta$
is labelled by a single integer, the degree of the curve in
$\IP^2$. By using the localization techniques of Kontsevich,
adapted to the noncompact case, one finds \cite{ckyz,kz}: \ben
\label{localp2fs} F_0 (q)&=& -{t^3 \over 18} + 3 \,q -{45 \,q^2
\over 8} + {244 \,q^3 \over 9} -
{12333 \, q^4 \over 64}  \cdots \nonumber\\
F_1 (q)&=& -{t \over 12} + {q \over 4} -{3 \, q^2 \over 8} -{23 \, q^3
\over 3} + {3437 \, q^4 \over 16} \cdots\nonumber\\
F_2 (q)&=& { \chi (X) \over 5720} + {q \over 80} + {3 \, q^3 \over
20} -{514 \, q^4 \over 5} \cdots \een and so on. In
(\ref{localp2fs}), $t$ is the K\"ahler class of the manifold,
$\chi(X)=2$ is the Euler characteristic of the local $\IP^2$, and
$q={\rm e}^{-t}$ . The first term in $F_2$ is the contribution of
constant maps, and we will provide later on a universal expression
for it.

It should be mentioned that there is of course a very powerful method to
compute $F_g$, namely mirror symmetry (the B-model). In the B-model, the
$F_g$ amplitudes are deeply related to the variation of complex structures on the
Calabi-Yau manifold (Kodaira-Spencer theory) and can be computed through
the holomorphic anomaly equations of \cite{bcov}. B-model computations of
Gromov-Witten invariants and $F_g$ amplitudes
can be found for example in \cite{bcov,ckyz,hosono,
kz,kkv}. Finally,
it should be mentioned that, when type II theory
is compactified on a Calabi-Yau
manifold, the $F_g$ appear naturally as the couplings of some special
set of F-terms of the low-energy supergravity action \cite{bcov,agnt}. This
point of view has shown to be extremely important in understanding the
properties of topological strings.

\subsection{Open topological strings}

Let us now consider open topological strings. The natural starting point
is a topological sigma model in which the worldsheet is now a Riemann
surface $\Sigma_{g,h}$ of genus $g$ with $h$ holes. Such models were
analyzed in detail in
\cite{csts}. The main issue is of course to specify boundary conditions for
the maps $x: \Sigma_{g,h} \rightarrow X$. It turns out that, for the
A-model, the relevant boundary conditions are Dirichlet, supported on
Lagrangian submanifolds of the Calabi-Yau $X$. If we denote by $C_i$, $i=1,
\cdots, h$ the holes of $\Sigma_{g,h}$ ({\it i.e.} the disconnected
components of the boundary $\partial \Sigma_{g,h}$), we have to pick
Lagrangian submanifolds ${\cal L}_i$, and consider maps such that
\begin{equation}
x(C_i)\subset {\cal L}_i.
\label{bound}
\end{equation}
These boundary
conditions are a consequence of requiring $Q$-invariance at the
boundary. One also has boundary conditions on the fermionic fields of the
theory, which require that $\chi$ and $\psi$ at the boundary $C_i$
take values on $x^*(T{\cal L}_i)$. We can also couple the theory to
Chan-Paton degrees of freedom on the boundaries,
giving rise to a $\otimes_i U(N_i)$ gauge symmetry.
The model can then be interpreted as a topological open string theory in the
presence of $N_i$ topological D-branes wrapping the Lagrangian submanifolds
${\cal L}_i$. Notice that, in contrast to physical D-branes in
Calabi-Yau manifolds,
which wrap special Lagrangian submanifolds \cite{bbs,ooy}, in the
topological framework the conditions are relaxed to just Lagrangian.

Once boundary conditions have been specified, one can define the
free energy of the topological string theory similarly to what we
did in the closed case. Let us consider for simplicity the case in
which one has a single Lagrangian submanifold ${\cal L}$, so that
all the boundaries of $\Sigma_{g,h}$ are mapped to ${\cal L}$.
Now, in order to specify the topological sector of the map, we
have to give two different kinds of data: the boundary part and
the bulk part. For the bulk part, the topological sector is
labelled by relative homology classes, since we are requiring the
boundaries of $x_*[\Sigma_{g,h}]$ to end on ${\cal L}$. Therefore,
we will set
\begin{equation}
x_*[\Sigma_{g,h}]={\cal Q}, \,\,\,\,\,\,\, {\cal Q}\in H_2(X, {\cal L})
\label{bulkpart}
\end{equation}
To specify the topological sector of the boundary, we will assume that
$b_1 ({\cal L})=1$, so that $H_1 ({\cal L})$ is generated by a nontrivial
one cycle $\gamma$. We then have
\begin{equation}
x_*[C_i]=w_i \gamma, \,\,\,\,\, w_i \in {\bf Z},\,\,\,\,
i=1, \cdots, h,
\label{wind}
\end{equation}
in other words, $w_i$ is the winding number associated to the map $x$
restricted to $C_i$. We will collect these integers into a single vector
$h$-uple denoted by $w=(w_1, \cdots, w_h)$.

There are various generating functionals that we can consider, depending on
the topological data that we want to keep fixed. It is very useful to fix
$g,h$ and the winding numbers, and sum over all bulk classes. This produces
the following generating
functional of open Gromov-Witten
invariants:
\begin{equation}
F_{w,g } (t) =\sum_Q F_{w,g}^{Q} {\rm e}^{-Q\cdot t}.
\end{equation}
In this equation, we have labelled the relative cohomology classes
${\cal Q}$ of embedded Riemann surfaces by a vector $Q$ of
$b_2(X)$ integers defined as
\begin{equation}
\int_{\cal Q} \omega =Q \cdot t,
\end{equation}
where $t=(t_1, \cdots, t_{b_2(X)})$ are the complexified K\"ahler
parameters of the Calabi-Yau manifold. In many examples relevant
to knot theory, the entries $Q$ are naturally chosen to be
half-integers. Finally, the quantities $F_{w,g}^{Q}$ are the open
string Gromov-Witten invariants, and they ``count" in an
appropriate sense the number of holomorphically embedded Riemann
surfaces of genus $g$ in $X$ with Lagrangian boundary conditions
specified by ${\cal L}$ and in the class represented by $Q,w$.
These are in general rational numbers.

We can now consider the total free energy, which is
the generating functional for all topological
sectors:
\begin{equation}
F(V)=\sum_{g=0}^{\infty} \sum_{h=1}^{\infty}
\sum_{w_1, \cdots, w_h} {i^h \over h!}
g_s^{2g-2+h} F_{g, w} (t)
{\rm Tr}\,V^{w_1} \cdots {\rm Tr}\, V^{w_h},
\label{totalfreeopen}
\end{equation}
where $g_s$ is the string coupling constant, and $V$ is a matrix source
that keeps track of the topological sector at the boundary.
The factor $i^h$ is very convenient
in order to compare to the Chern-Simons free energy, as we will see
later. The factor $h!$ is a symmetry factor which takes into
account that the holes are indistinguishable (or one could have
absorbed them into the definition of $F_{g,w}$).

In order to compare open Gromov-Witten invariants to knot invariants,
it is useful to introduce the following notation.
When all $w_i$ are positive, one can label $w$
in terms of a vector $\vec k$. Given an $h$-uple $w=(w_1, \cdots, w_h)$,
we define a vector $\vec k$ as follows: the $i$-th
entry of $\vec k$ is the number of $w_j$'s
which take the value $i$. For example, if $w_1=w_2=1$ and
$w_3=2$, this corresponds to $\vec k =(2,1,0,\cdots)$.
In terms of $\vec k$, the number of holes and the total winding number are
\begin{equation}
h=|\vec k|\equiv \sum_j k_j,\,\,\,\ \ell =\sum_i w_i=\sum_j  j k_j.
\end{equation}
Note that a given $\vec k$ will correspond to many $w$'s
which differ by permutation of entries.  In fact
there are $h!/\prod_j k_j!$ $h$-tuples $w$ which
give the same vector $\vec k$ (and the same amplitude). We can
then write the total free energy for positive winding numbers as:
\begin{equation}
F(V)=\sum_{g=0}^{\infty} \sum_{\vec k} {i^{|\vec k|} \over \prod_j k_j!}
g_s^{2g-2+h} F_{g, \vec k} (t) \Upsilon_{\vec k} (V)
\end{equation}
where
\begin{equation}
\Upsilon_{\vec k} (V)=\prod_{j=1}^{\infty} ({\rm Tr} V^j)^{k_j}.
\label{ups}
\end{equation}

Although a rigorous theory of open Gromov-Witten invariants is not
available, localization techniques make possible to compute them in various
situations \cite{kl,song,gz,mayr,blum,klemm}. It is also possible to use mirror
symmetry to compute disc invariants ({\it i.e.} when $g=0$, $h=1$), as it
was first shown in \cite{av} and subsequently explored in
\cite{akv,mayrone,lema,iqbaldisc,sarkar}. Finally, we also mention that the open string
amplitudes $F_{g,w}$ also appear as low-energy couplings of type II superstrings compactified
on Calabi-Yau manifolds in the presence of D-branes \cite{bcov,vafa}.

\subsection{Integer invariants from topological strings}

The closed and open Gromov-Witten invariants that have been introduced are
both rational, due to the orbifold structure of the moduli spaces. On the other
hand, these invariants are deeply related to questions in
enumerative geometry, but the
relation between the invariants and the number of holomorphic curves of a given
genus and in a given homology class is far from being simple.
An obvious reason for this is {\it multicovering}. Suppose
you have found a holomorphic map $x: {\bf S}^2 \rightarrow X$ in genus zero
of degree $d$. Then, simply by composing this with a degree $k$ cover ${\bf S}^2
\rightarrow {\bf S}^2$, you get another holomorphic map of degree
$kd$. Therefore, at every degree, in order to count the actual number
of ``primitive" holomorphic curves, one should subtract the
contributions coming from
multicovering of curves with lower degree. On top of that, the contribution
of a $k$-cover appears in $N_{0,kd}$ with weight $k^{-3}$. Therefore,
although in genus zero the Gromov-Witten invariants are not integer, this is due
to the effects of multicovering, and once this has been taken into account
one extracts integer numbers that correspond in many cases to actual numbers of
rational curves. The multicovering phenomenon at genus $0$ was found experimentally
in \cite{cdgp} and later on derived in \cite{am}.

Another geometric effect that has to be taken into account is
bubbling \cite{bcovhol,bcov}. Imagine that you found a
map $x: \Sigma_g \rightarrow X$ from a
genus $g$ surface to a Calabi-Yau threefold. By gluing to $\Sigma_g$ a
small Riemann surface of genus $h$, and making it very small, you
get an approximate holomorphic map from a Riemann surface whose genus is
topologically $g+h$. This means that ``primitive'' maps at
genus $g$ contribute to all genera $g'>g$, and in order to count curves
properly we should take this effect into account.

These facts suggest that, although the Gromov-Witten invariants are not
in general integer numbers, they have some hidden integrality structure, and
that one can extract from them integer invariants that are related to a counting
problem. But it turns out that, instead of deriving
the various effects of multicovering and bubbling
from a geometrical point of view, the underlying integral structure of the
Gromov-Witten invariants is better revealed when the $F_g$ is
regarded as a low-energy coupling in a
compactification of type IIA theory on a
Calabi-Yau manifold. Using this approach, Gopakumar and Vafa showed
\cite{gv} that
one can write the generating functional $F(g_s, t)$ in terms of contributions
associated to BPS states,
and they used type IIA/M-theory duality to obtain a completely new
point of view on topological strings. They showed in particular
that Gromov-Witten invariants of closed strings
can be written in terms
of some new, {\it integer}
invariants known as {\it Gopakumar-Vafa invariants}.
These invariants count in a very precise way the number of
BPS states that arise
in the Calabi-Yau compactification of type IIA theory.
We will now describe this result in some detail and provide some examples.

The result of Gopakumar and Vafa concerns the overall structure of
$F(g_s,t)$. According to \cite{gv}, the generating functional (\ref{freen})
can be written as
\begin{equation}
F(g_s,t)=\sum_{g=0}^{\infty} \sum_{\beta} \sum_{d=1}^{\infty}
n^g_{\beta} {1\over d}\biggl( 2 \sin {d g_s \over 2} \biggr)^{2g-2} q^{d\beta},
\label{gvseries}
\end{equation}
where $n^g_\beta$, which are the Gopakumar-Vafa invariants, are
integer numbers. In (\ref{gvseries}), $t$ denotes the set of
$b_2(X)$ K\"ahler parameters, and $q^{\beta}$ is defined as in
(\ref{threepoint}). It is very illuminating to expand
(\ref{gvseries}) in powers of $g_s$ and extract from it the
structure of a given $F_g$. One easily obtains, for $g=0$, the
well-known structure of the prepotential \cite{cdgp,am}: \be
F_0={1\over 3!}\int_X \omega^3 + \int_X c_2 (X) \wedge \omega +
\chi (X) {\zeta(3) \over 2}+ \sum_{\beta} n_{\beta}^0 {\rm Li}_3
(q^{\beta}), \ee up to the polynomial terms in $t$. Here
$\chi(X)$, $c_2 (X)$ denote respectively the Euler characteristic
and the second Chern class of the Calabi-Yau target. We recall
that ${\rm Li}_j$ denotes the polylogarithm of index $j$, which is
defined by:
\begin{equation}
{\rm Li}_j (x)= \sum_{n=1}^{\infty} {x^n \over n^j}.
\end{equation}
Notice that ${\rm Li}_1 (x) =-\log (1-x)$, while for $j\le 0$, ${\rm
Li}_j(x)$ is a rational function of $x$:
\be
{\rm Li}_j(x)= \Bigl( x {d \over dx}\Bigr)^{|j|} {1 \over 1-x} =
|j|! { x^{|j|} \over (1-x)^{|j|+1}} +\cdots.
\ee
For $g=1$, one obtains:
\be
\label{elliptic}
F_1={1 \over 24} \int_X c_2 (X) \wedge \omega  +
\sum_{\beta} \Bigl( {1 \over 12} n_{\beta}^0 + n_{\beta}^1 \Bigr)
{\rm Li}_1 (q^{\beta}).
\ee
Finally, for $g>1$, the Gopakumar-Vafa result gives:
\ben
F_g (t) &=& {(-1)^g \chi(X) |B_{2g} B_{2g-2}| \over
4g (2g-2)(2g-2)!} \nonumber\\
&+&
\sum_{\beta} \biggl( { |B_{2g}| n_{\beta}^0 \over
2g (2g-2)!} + {2 (-1)^g n_{\beta}^2 \over (2g-2)!}
\pm \cdots - {g-2 \over 12} n^{g-1}_{\beta} + n_{\beta}^g\biggr) {\rm Li}_{3-2g}(q^{\beta}).
\label{multibuble}
\een
In this equation, $B_n$ denote the Bernoulli numbers.
The first term in (\ref{multibuble}) is the contribution to $F_g$
associated to maps from $\Sigma_g$ to a single point. Comparing
it with (\ref{hint}) we find that the Gopakumar-Vafa structure result
predicts:
\be
\int_{{\overline M}_g} c^3_{g-1}(\DE)= {|B_{2g} B_{2g-2}| \over
2g (2g-2)(2g-2)!}.
\ee
This expression was conjectured by Faber \cite{fabercon},
derived in \cite{mm} from heterotic/type IIA duality, and proved in
\cite{fp}.

The polylogarithm in (\ref{multibuble}) indicates that the degree
$k$ multicover of a curve of genus $g$ contributes with a factor
$k^{2g-3}$ to $F_g$.
This generalizes the results of \cite{cdgp} for genus $0$
and results for genus $1$ in
\cite{bcovhol}. The multicover contribution was also found in \cite{mm} by
using heterotic/type II duality. But equation (\ref{multibuble}) also takes
into account in a precise way the effect of bubbling on $F_g$:
at every genus $g$,
one has to take into account all the previous genera $g'<g$ in order to
extract the Gopakumar-Vafa invariants $n_\beta^g$.

The Gopakumar-Vafa invariants contain all the information of the
Gromov-Witten invariants, and vice versa: if one knows the
Gopakumar-Vafa invariants $n_\beta^g$ for all $g$ and $\beta$, one
can deduce the $N_{g,\beta}$, and the other way around. This
follows just by comparing (\ref{gvseries}) with (\ref{freen}), and
it is worked out in detail in \cite{bp}, where explicit formulae
for the relation between $N_{g,\beta}$ and $n^g_{\beta}$ are
given. But one remarkable aspect of the Gopakumar-Vafa picture is
that, in many situations, the integer invariants $n_{\beta}^g$ can
be computed much more easily than their Gromov-Witten counterparts
\cite{gv,kkv}. In fact, their computation involves in many cases
just classical algebraic geometry, so one gets rid of the
complications of the moduli space of maps. The physical reason
behind is that in the Gopakumar-Vafa picture one looks at
worldsheet instantons using the physical gauge approach (in the
terminology of \cite{wscorr}), {\it i.e.} one views the worldsheet
instanton as a submanifold of the target, and not as a map
embedding a Riemann surface $\Sigma_g$ inside a Calabi-Yau.
Related developments can be found in \cite{hk}.

Let us consider some simple examples of the Gopakumar-Vafa
invariants. The simplest one refers to the noncompact Calabi-Yau manifold
${\cal O}(-1) \oplus {\cal O}(-1) \rightarrow \IP^1$, also known as the
resolved conifold, which will play an important role later on. This
manifold is toric, and can be described as the zero locus of
\be
|x_1|^2 + |x_4|^2 -|x_2|^2- |x_3|^2=s
\label{toricres}
\ee
quotiented by a $U(1)$ that acts as
\be
x_1, x_2, x_3, x_4 \rightarrow {\rm e}^{i \alpha} x_1, {\rm e}^{-i \alpha}
x_2, {\rm e}^{-i \alpha} x_3,
 {\rm e}^{i \alpha} x_4
\ee This is the description that appears naturally in the linear
sigma model of \cite{phases}. Notice that, for $x_2=x_3=0$,
(\ref{toricres}) describes a $\IP^1$ whose area is proportional to
$s$. Therefore, $(x_1, x_4)$ can be taken as homogeneous
coordinates of the $\IP^1$ which is the basis of the fibration,
while $x_2, x_3$ can be regarded as coordinates for the fibers.
This manifold has $b_2(X)=1$, corresponding to the $\IP^1$ in the
base, and its total free energy turns out to be \be
F(g_s,t)=\sum_{d=1}^{\infty} {1\over d \Bigl(2 \sin {d g_s \over
2} \Bigr)^{2}} q^{d}, \label{resf}
\end{equation}
where $q={\rm e}^{-t}$ and $t$ is the complexified area of the
$\IP^1$. We see that the only nonzero Gopakumar-Vafa invariant is
$n_1^0=1$. On the other hand, this model already has an infinite
number of nontrivial $N_{g,\beta}$ invariants, but these are all
due to bubbling and multicovering: the model only has one true
``primitive'' curve, which is just $\IP^1$, and this is what the
Gopakumar-Vafa invariant is computing.

A more complicated example is the local $\IP^2$ geometry
considered before, which already has an infinite number of
nontrivial Gopakumar-Vafa invariants. These have been computed in
\cite{kz,kkv,amv} using the A-model, the B-model, and the duality
with Chern-Simons theory that we will explain in section 6. Some
results are presented in Table \ref{Pgv}. In this table, the
integer $d$ labels the class $\beta$, and corresponds to the
degree of the curve in $\IP^2$. Notice that the first
Gromov-Witten invariants are $N_{0,1}=3$, and $N_{0,2}=-45/8$, as
listed in (\ref{localp2fs}), therefore using the
multicovering/bubbling formula one finds $n_1^0=N_{0,1}=3$, and
$N_{0,2}=n_1^0/8 + n_2^0$, which gives $n_2^0=-6$.
\begin{table}
\begin{center}
\begin{tabular}{|cccccc|}
 \hline
 $d$&$g=0$ & 1&2&3&4 \\  \hline
 1&3&0&0&0&0\\ \hline
 2&-6&0&0&0&0\\ \hline
3&27&-10&0&0&0\\ \hline
4& -192 & 231 & -102 & 15 & 0 \\ \hline
5& 1695 &-4452 & 5430 & -3672 & 1386 \\ \hline
\end{tabular}
\end{center}
\caption{Gopakumar-Vafa invariants
$n_d^g$ for ${\cal O}(-3) \rightarrow \IP^2$.}
\label{Pgv}
\end{table}

For open topological strings one can derive a similar expression
relating open Gromov-Witten invariants to a new set of integer
invariants, that we will denote by $n_{w,g,Q}$. The corresponding
multicovering/bubbling formula was derived in \cite{ov,lmv},
following arguments similar to those in \cite{gv}, and states that
the free energies of open topological string theory in the sector
labelled by $w$ can be written in terms of the integer invariants
$n_{w, g, Q}$ as follows:
\begin{eqnarray}
& &\sum_{g=0}^{\infty}
g_s^{2g-2 +h}
F_{w,g}(t) = \nonumber\\& &{1\over \prod_i w_i }
\sum_{g=0}^{\infty}\sum_{d|w}  (-1)^{h+g}\,
n_{w/d, g, Q}\, d^{h-1}
\biggl( 2\sin {d g_s \over 2} \biggr)^{2g-2}
\prod_i \biggl( 2\sin {w_i g_s \over 2} \biggr) {\rm e}^{-d Q \cdot t}.
\label{multopen}
\end{eqnarray}
Notice there is one such identity for each $w$.
In this expression, the sum is over all integers $d$ which satisfy that
$d|w_i$ for all $i=1, \cdots, h$.
When this is the case, we define the $h$-uple $w/d$
whose $i$-th component is $w_i/d$. The expression (\ref{multopen})
can be expanded to give a set of multicovering/bubbling formulae for
different genera. Up to genus $2$ one finds,
\begin{eqnarray}
\label{multiopen}
F_{w, g=0}^{Q}&=& (-1)^{h}
\sum_{d|w} d^{h-3} n_{w/d , 0, Q/d} ,\nonumber\\
F_{w ,g=1}^{Q}&=&-(-1)^{h}
\sum_{d|w} \biggl( d^{h -1} n_{w/d,1, Q/d}-
{d^{h -3} \over 24}\bigl(2d^2-\sum_i w_i^2 \bigr)\,
n_{w/d,0, Q/d}\biggr) ,\cr
F_{\vec k, g=2}^{Q}&=&(-1)^{h}
\sum_{d|w} \biggl(d^{h+1}   n_{w/d, 2, Q/d} +
{d^{h-1}   \over 24 }n_{w/d, 1, Q/d} \sum_i w_i^2
 \nonumber\\ &
+& {d^{h - 3} \over 5760}
\bigl(24 d^4 - 20 d^2 \sum_i w_i^2 -2\sum_i w_i^4 +
5\sum_{i_1, i_2} w^2_{i_1} w^2_{i_2} \bigr)\,
n_{w/d, 0, Q/d} \biggr).
\end{eqnarray}
In these equations, the integer $d$ has to divide the vector $w$ (in the
sense explained above) and it is understood that $n_{w_d,g,Q/d}$
is zero if $Q/d$ is not a relative homology
class.

It is important to notice that the integer invariants $n_{w, g,
Q}$ are not the most fundamental ones. When all the winding
numbers are positive, we can represent $w$ by a vector $\vec
k=(k_1, k_2, \cdots)$, as we explained in 2.3. Such a vector can
be interpreted as a label for a conjugacy class $C(\vec k)$ of the
symmetric group $S_{\ell}$, where $\ell=\sum_j j k_j$ is the total
winding number: $C(\vec k)$ is the conjugacy class with $k_1$
one-cycles, $k_2$ two-cycles, and so on. The invariant $n_{w, g,
Q}$ will be denoted as $n_{\vec k, g, Q}$, and D-brane physics
states that it can be written as \be n_{\vec k, g, Q}=\sum_R
\chi_R (C(\vec k)) N_{R, g, Q}, \label{openBPS} \ee where $ N_{R,
g, Q}$ are integer numbers labelled by representations of the
symmetric group, {\it i.e.} by Young tableaux, and $\chi_R$ is the
character of $S_{\ell}$ in the representation $R$. The above
relation is invertible, since by orthonormality of the characters
one has \be N_{R, g, Q}=\sum_{\vec k} { \chi_R (C(\vec k)) \over
z_{\vec k}} n_{\vec k, q, Q}, \ee where
\begin{equation}
z_{\vec k}={\ell!\over  |C(\vec k)|}= \prod k_j! \prod j^{k_j}.
\label{zk}
\end{equation}
Notice that integrality of $N_{R,g,Q}$ implies integrality of $n_{\vec k,
q, Q}$, but not the other way around. In that sense, the invariants $N_{R,
g, Q}$ are more fundamental. We will further clarify this issue in section 4.

\sectiono{Chern-Simons theory and knot invariants}

In this section we make a short review of Chern-Simons theory and
its relations to knot invariants.

\subsection{Chern-Simons theory: basic ingredients}

Chern-Simons theory, introduced by Witten in \cite{cs}, provides a quantum
field theory description of a wide class of invariants of three-manifolds
and of knots and links in three-manifolds. The Chern-Simons
action with gauge group $G$
on a generic three-manifold $M$ is defined by
\begin{equation}
S={k \over 4\pi} \int_M {\rm Tr} \Bigl( A\wedge d A + {2 \over 3} A
\wedge A \wedge A \Bigr).
\label{csact}
\end{equation}
Here, $k$ is the coupling constant, and $A$ is a $G$-gauge connection on the
trivial bundle over $M$. We will assume for simplicity that $G$ is a
simply-laced group, unless otherwise stated. As noticed by Witten,
since this action does not involve the metric, the resulting
quantum theory is topological, at least formally. In particular,
the partition function
\begin{equation}
Z_k(M)= \int [{\cal D} A]  {\rm e}^{iS}
\label{partcs}
\end{equation}
should define a topological invariant of the manifold $M$. A
detailed analysis \cite{cs} shows that this is in fact the case,
with an extra subtlety: the invariant depends on the
three-manifold {\it and} of a choice of framing, {\it i.e.} a
choice of trivialization of the bundle $TM \oplus TM$ (this should
be called, strictly speaking, a 2-framing, but we will refer to it
as framing, following standard practice). As explained in
\cite{atiyah}, for every three-manifold there is a canonical
choice of framing, and the different choices are labelled by an
integer $s \in {\bf Z}$ in such a way that $s=0$ corresponds to
the canonical framing. In the following all the results will be
presented in the canonical framing.

The partition function of Chern-Simons theory can be computed in a variety
of ways. One can for example use perturbation theory and produce an
asymptotic series in $k$ around a classical solution to the action.
The classical solutions of
Chern-Simons theory are just flat connections $F(A)=0$
on $M$. Let us assume that
these are a discrete set of points (this happens, for example, if $M$ is a
rational homology sphere). In that situation, one expresses $Z_k(M)$ as a
sum
of terms associated to stationary points:
\begin{equation}
Z_k (M)= \sum_c Z_k^{(c)}(M),
\end{equation}
where $c$ labels the different flat connections $A^{(c)}$ on $M$.
The structure of the perturbative series was analyzed in
various papers \cite{cs,roz,as} and is given by the following
expression:
\begin{equation}
\label{perts}
Z_k^{(c)}(M)=Z^{(c)}_{\rm 1-loop}(M). \exp \Biggl\{ \sum_{\ell=1}^\infty
S^{(c)}_\ell  x^\ell
\Biggr\}.
\end{equation}
In this equation, $x$ is the effective expansion parameter:
\begin{equation}
\label{coupling}
x = { 2 \pi i \over k+y},
\end{equation}
where $y$ is the dual Coxeter of the group, and we will set $l=k
+y$. For $G=SU(N)$, $y=N$. The one-loop correction $Z^{(c)}_{\rm
1-loop}(M)$ was first analyzed in \cite{cs}, and studied in
great detail since then. It 
involves some important normalization factors of the
path-integral, and determinants of differential
operators. After some work it can be written in terms of 
topological invariants of the three-manifold and 
the flat connection $A^{(c)}$,
\begin{equation}
\label{asympt} Z^{(c)}_{\rm 1-loop}(M)= { (2 \pi x)^{ {1\over
2}({\rm dim}\, H^0_{A^{(c)}} - {\rm dim}\, H^1_{A^{(c)}} )} \over
{\rm vol} (H_c)} {\rm e}^{-{1 \over x} S_{\rm CS}(A^{(c)}) -{i \pi
\over 4}\varphi} {\sqrt {| \tau^{(c)}_R|}},
\end{equation}
where $H^{0,1}_{A^{(c)}}$ are the de Rham cohomology groups with
values in the Lie algebra of $G$ and associated to the trivial
connection $A^{(c)}$, $ \tau^{(c)}_R$ is the
Reidemeister-Ray-Singer torsion of $A^{(c)}$, $H_c$ is the
isotropy group of $A^{(c)}$, and $\varphi$ is a certain phase. 
More details about the structure of this term can be found in
\cite{cs,fg,jeffrey,roz}. The terms $S^{(c)}_\ell$ in
(\ref{perts}) correspond to connected diagrams at $\ell+1 $ loops,
and since they involve evaluation of group factors of Feynman
diagrams, they depend explicitly on the gauge group $G$ and the
isotropy subgroup $H_c$. In the $SU(N)$ or $U(N)$ case, and for
$A^{(c)}=0$ (the trivial flat connection) they are polynomials in
$N$. For the trivial flat connection, one also has that ${\rm
dim}\, H^0_{A^{(c)}}={\rm dim}\, G$, ${\rm dim}H^1_{A^{(c)}}=0$,
and $H_c=G$. The terms $S^{(c)}_\ell$ are also topological invariants
associated to the 
three-manifold and the flat connection, and they emerge naturally from the
perturbative analysis of Chern-Simons theory.   

As Witten showed in \cite{cs}, it is also
possible to use nonperturbative methods to obtain a combinatorial formula
for (\ref{partcs}). This goes as follows.
\begin{figure}
\leavevmode
\begin{center}
\epsfysize=5cm
\epsfbox{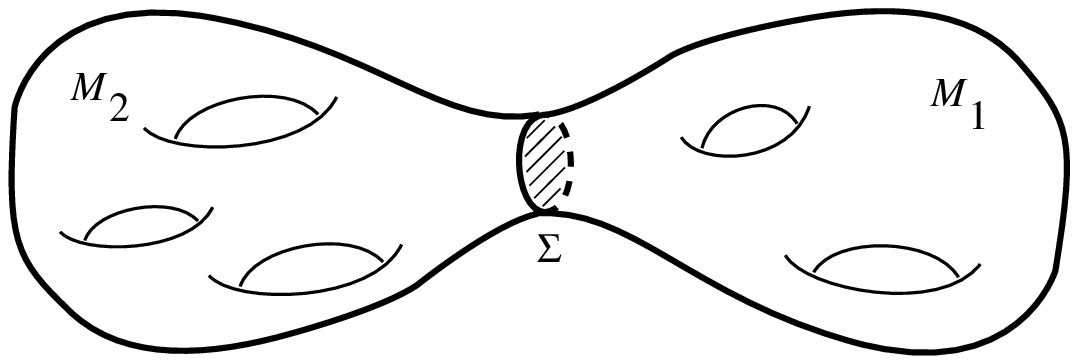}
\end{center}
\caption{Heegard splitting of a three-manifold $M$ into two three manifolds
$M_1$ and $M_2$ with a common boundary $\Sigma$.}
\label{heegard}
\end{figure}
By canonical quantization, one associates a Hilbert
space ${\cal H}(\Sigma)$ to any two-dimensional compact manifold that arises as the boundary
of a three-manifold, so that the path-integral over a manifold with
boundary gives a state in the corresponding Hilbert space. In order to
compute the partition function of a three-manifold $M$, one can perform a
Heegard splitting {\it i.e.} represent $M$ as the
connected sum of two three-manifolds $M_1$ and $M_2$ sharing a common
boundary $\Sigma$, where $\Sigma$ is a Riemann surface. If $f: \Sigma
\rightarrow \Sigma$ is a homeomorphism, we will write $M=M_1 \cup_f
M_2$, so that $M$ is obtained by gluing $M_1$ to $M_2$ through their common
boundary by using the homeomorphism $f$. This is represented in
\figref{heegard}. We can then compute the full path integral (\ref{partcs})
over $M$ by
computing first the path integral over $M_1$ and $M_2$. This produces two
wavefunctions $|\Psi_{M_1}\rangle$, $|\Psi_{M_2}\rangle$ in ${\cal H} (\Sigma)$.
On the other
hand, the homeomorphism $f: \Sigma \rightarrow \Sigma$ will be
represented by an operator in the Hilbert space,
\be
U_f: {\cal H}(\Sigma) \rightarrow {\cal H}(\Sigma)
\ee
and the partition function can then be evaluated as
\be
Z_k(M)=\langle \Psi_{M_2}| U_f | \Psi_{M_1}\rangle.
\ee
In order to use this method, we have to find first the Hilbert space
associated to a boundary. There is one special case in which this can be
done quite systematically, namely when $\Sigma={\bf T}^2$, a two-torus.
As it was first shown
in \cite{cs} (and worked out in detail in \cite{torus,lr,knotops}),
the states of the Hilbert space of
Chern-Simons theory associated to the torus, ${\cal H} ({\bf T}^2)$,
 are in one
to one correspondence with
the integrable representations of the Wess-Zumino-Witten (WZW) model
with gauge group $G$ at
level $k$ \footnote{We will use the following notations in the following:
the fundamental weights of $G$ will be denoted by
$\lambda_i$, the simple roots by $\alpha_i$, with $i=1, \cdots, r$,
and $r$ denotes the rank of $G$.
The weight and root lattices of $G$ are
denoted by $\Lambda_{\rm w}$ and $\Lambda_{\rm r}$, respectively, and
$|\Delta_+|$ denotes the number of positive roots.}.
A representation given by a highest weight $\Lambda$ is integrable if
the weight $\rho + \Lambda$ is in the
Weyl alcove ${\cal F}_l$, where $l=k+y$ and $\rho$ denotes as usual the Weyl vector,
given by the sum of the fundamental weights. The Weyl alcove
is given by $\Lambda_{\rm w}/l
\Lambda_{\rm r}$ modded out by the action of the Weyl group. For example,
in $SU(N)$ a
weight $p=\sum_{i=1}^r p_i \lambda_i$ is in ${\cal F}_l$ if
\begin{equation}
\sum_{i=1}^r p_i < l,\,\,\,\,\,\, {\rm and} \,\,\ p_i >0, \, i=1, \cdots, r.
\end{equation}
In the following, the basis of integrable representations will be
labelled by the weights in ${\cal F}_l$, and the states in the
Hilbert state of the torus ${\cal H}({\bf T}^2)$ will be denoted
by $ |p\rangle = | \rho + \Lambda \rangle$ where $\Lambda$, as we
have stated, is an integrable representation of the WZW model at
level $l$. The states $|p\rangle$ can be chosen to be orthonormal
\cite{torus,lr,knotops}.

There is a special class of homeomorphisms of ${\bf T}^2$ that have a
simple expression as operators in ${\cal H}({\bf T}^2)$. These are ${\rm
Sl}(2, {\bf Z})$ transformations, whose generators $T$ and $S$ have the
following simple matrix elements in the above basis:
\ben
\label{st}
T_{\alpha \beta}&=& \delta_{\alpha\beta}
 {\rm e}^{2\pi i (h_\alpha -c/24)},\nonumber\\
S_{\alpha\beta}&=& {i^{|\Delta_+|} \over (k+y)^{r/2}} \Biggl( {{\rm Vol} \,
\Lambda^w \over{\rm Vol} \,
\Lambda^r} \Biggr)^{1\over 2} \sum_{w \in {\cal W}} \epsilon (w) \exp \Bigl(
-{2 \pi i \over k+y} \alpha \cdot w(\beta)\Bigr).
\een
In the first equation, $h_\alpha$ is the conformal weight of the primary field
associated to $\alpha$:
\be
h_\alpha = {\alpha^2 -\rho^2 \over 2(k+y)},
\label{confweight}
\ee
and $c$ is the central charge of the WZW model.
In the second equation, the sum over $w$ is a sum over the elements of the
Weyl group ${\cal W}$, and $\epsilon (w)$ is the signature of $w$.
These explicit formulae allow us to compute the partition function of
any three-manifold that
admits a Heegard splitting along a torus, like for example a lens space. The
case of ${\bf S}^3$ is particularly simple. It is well-known that ${\bf
S}^3$ can be obtained by gluing two solid tori along their boundaries
through an $S$ transformation. The wavefunction associated to the solid
torus is simply the vacuum, which corresponds to $|\rho\rangle$, and we find
\begin{equation}
Z({\bf S}^3)=\langle \rho |S | \rho \rangle = S_{\rho \rho}.
\label{pfsphere}
\end{equation}
By using Weyl's denominator formula,
\be
\label{wdf}
\sum_{w \in {\cal W}} \epsilon (w) {\rm e}^{w(\rho)} =\prod_{\alpha>0}
2 \sinh {\alpha \over 2},
\ee
one finds
\be
Z({\bf S}^3)= {1 \over (k+y)^{r/2}} \Biggl( {{\rm Vol} \,
\Lambda^w \over{\rm Vol} \,
\Lambda^r} \Biggr)^{1\over 2} \prod_{\alpha>0}
2 \sin \Bigl( {\pi (\alpha \cdot \rho) \over k+y} \Bigr).
\label{css3}
\ee

\begin{figure}
\leavevmode
\begin{center}
\epsfysize=13cm
\epsfbox{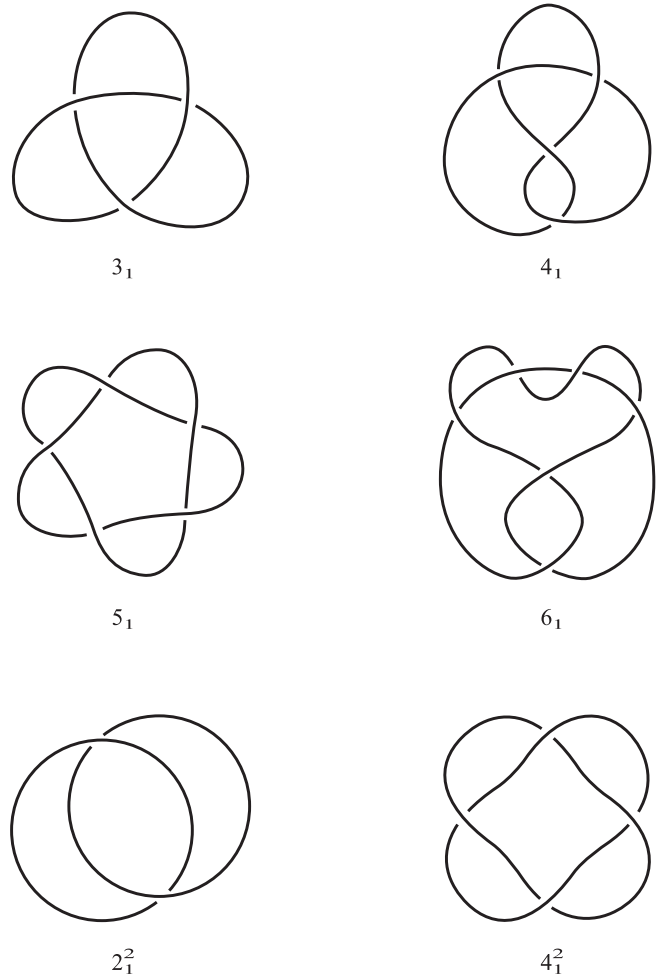}
\end{center}
\caption{Some knots and links. In the notation $x_n^L$, $x$
indicates the number of crossings, $L$ the number of components (in case it
is a link with $L>1$) and $n$ is a number used to enumerate knots and links
in a given set characterized by $x$ and $L$. The knot $3_1$ is also known
as the trefoil knot, while $4_1$ is known as the figure-eight knot. The
link $2_1^2$ is called the Hopf link. }
\label{nudos}
\end{figure}

Besides providing invariants of three-manifolds, Chern-Simons theory also
provides invariants of knots and links inside three-manifolds
(for a survey of modern knot theory, see \cite{lick,ps}). Some examples
of knots and links are depicted in \figref{nudos}.
When dealing with knots, we will always consider that the Chern-Simons
gauge group is $G=SU(N)$ or $U(N)$.
Given a knot ${\cal K}$ in ${\bf S}^3$, we can
consider the trace of the holonomy of the gauge connection around
$\cal K$ in a given irreducible representation $R$ of $SU(N)$, which gives
the Wilson loop operator:
\begin{equation}
W^{\cal K}_R(A)={\rm Tr}_R \Bigl( {\rm P}\,\exp\, \oint_{\gamma}
A\Bigr),  \label{wilson}
\end{equation}
where ${\rm P}$ denotes path-ordered exponential. This is a gauge
invariant operator whose definition does not involve the metric on
the three-manifold. The irreducible representations of $SU(N)$ can
be labelled by highest weights or equivalently by the lengths of
rows in a Young tableau, $l_i$, where $l_1 \ge l_2 \ge \cdots$. If
we now consider a link ${\cal L}$ with components ${\cal K}_i$,
$i=1, \cdots, L$, we can in principle compute the correlation
function,
\begin{equation}
W_{(R_1, \cdots, R_L)}({\cal L})=\langle W^{{\cal K}_1}_{R_1}\cdots
W^{{\cal K}_L}_{R_L}\rangle =
{1\over Z(M)}\int [{\cal D} A] \Bigl( \prod_{i=1}^L W_{R_i}^{{\cal K}_i}
\Bigr) {\rm e}^{iS}.
\label{vevknot}
\end{equation}
The topological character of the action, and the fact that the Wilson loop
operators can be defined without using any metric on the three-manifold,
indicate that (\ref{vevknot}) is a topological
invariant of the link ${\cal L}$. These correlation functions can be
studied in a variety of ways. The nonperturbative approach pioneered by
Witten in \cite{cs}, by exploiting the relation with WZW model, shows that
these correlation functions are rational functions of $q^{\pm {1\over 2}}$, $\lambda^{\pm
{1 \over 2}}$, where
\begin{equation}
q={\rm e}^x=\exp \Bigl( {2 \pi i \over k+N} \Bigr),\,\,\,\, \lambda =q^N.
\label{varias}
\end{equation}
It turns out that the correlation function (\ref{vevknot}) is the
quantum group invariant of the link ${\cal L}$ associated to the
 irreducible representations $R_1, \cdots, R_L$ of $U_q({\rm su}(N))$ 
(see for example \cite{rosso} for a general definition of the quantum group
invariant).

\begin{figure}
\begin{center}
\epsfxsize=5in\leavevmode\epsfbox{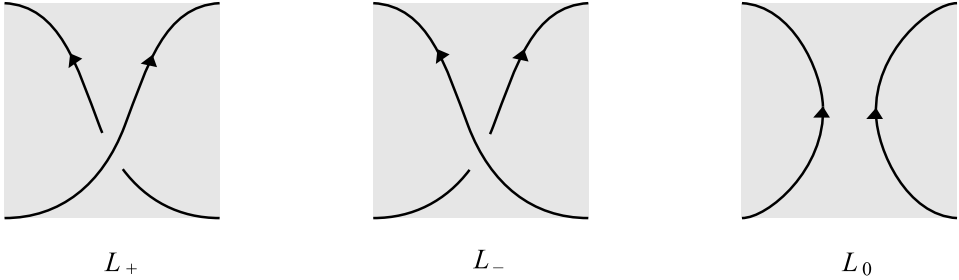}
\end{center}
\caption{Skein relations for the HOMFLY polynomial.}
\label{skein}
\end{figure}

The invariants of knots and links obtained as correlation functions
in Chern-Simons theory include and generalize
the HOMFLY polynomial \cite{homfly} (which is a generalization itself
of the Jones polynomial).
The HOMFLY polynomial of a link ${\cal L}$,
$P_{\cal L}(q, \lambda)$, can be defined through
the so-called {\it skein relation}. This goes as follows.
Let ${\cal L}$ be a link in ${\bf
S}^3$, and let us focus on one of the crossings in its plane projection. The crossing can be
an overcrossing, like the one depicted in $L_+$ in \figref{skein},
or an undercrossing, like
the one depicted in $L_-$. If the crossing is $L_+$,
we can form two other links either by undoing the crossing (and producing
$L_0$ of \figref{skein}) or by changing $L_+$ into $L_-$. In both cases the
rest of the link is left unchanged. Similarly, if the
crossing is $L_-$, we form two links by changing $L_-$ into $L_+$ or into
$L_0$. The links produced in this way will be in general topologically
inequivalent to the original one (they can even have a different number of
components). The skein relation
\begin{equation}
\lambda^{1 \over 2} P_{L_+} -\lambda^{-{1\over 2}}P_{L_-} =
(q^{1\over2} -q^{-{1\over 2}})P_{L_0}
\label{skeinrel}
\end{equation}
expresses the HOMFLY polynomial of the original link in terms of the links
that are obtained by changing the crossing. By using recursively this
relation, one can undo all the crossings and express the polynomial in
terms of its value on the unknot, or trivial knot. This value is usually
taken to be $P=1$. The HOMFLY polynomial corresponds
to a Chern-Simons $SU(N)$ link invariant with all the components in the
fundamental representation $R_{\alpha}=\tableau{1}$:
\begin{equation}
W_{(\tableau{1}, \cdots,\tableau{1})}({\cal L})=
\lambda^{{\rm lk}({\cal L})}
\biggl( {\lambda^{1\over 2} -\lambda^{-{1\over 2}}
\over q^{1\over 2} -q ^{-{1\over 2}}} \biggr) P_{\cal L}(q, \lambda)
\label{homflyrel}
\end{equation}
where ${\rm lk}({\cal L})$ is the linking number of ${\cal L}$. This can be
shown, as in \cite{cs}, by proving that the vev in the fundamental
representation satisfies the skein relation.

The link invariants defined in (\ref{vevknot}) can be computed in
many different ways. A particularly useful framework is the
formalism of knot operators \cite{knotops}. In this formalism, one
constructs operators that ``create'' knots wrapped around a
Riemann surface in the representation $R$ of the gauge group
associated to the highest weight $\Lambda$: \be W_{\Lambda}^{\cal
K}: {\cal H}(\Sigma) \rightarrow {\cal H}(\Sigma). \ee Notice that
the topology of $\Sigma$ restricts the type of knots that one can
consider. So far these operators have been constructed in the case
when $\Sigma={\bf T}^2$. The knots that can be put on a torus are
called {\it torus knots}, and they are labelled by two integers
$(n,m)$ that specify the number of times that they wrap the two
cycles of the torus. Here, $n$ refers to the winding number around
the noncontractible cycle of the solid torus, while $m$ refers to
the contractible one. The trefoil knot $3_1$ in \figref{nudos} is
the $(2,3)$ torus knot, and the knot $5_1$ is the $(2, 5)$ torus
knot. The operator that creates the $(n,m)$ torus knot will be
denoted by $W^{(n,m)}_{\Lambda}$, and it has a fairly explicit
expression: \be W_{\Lambda}^{(n,m)}|p\rangle = {\rm e}^{2 \pi i nm
h_{ \rho + \Lambda}} \sum_{\mu \in M_{\Lambda}}\exp \biggl[ -i\pi
\mu ^2 {nm \over k+N} - 2\pi i {m \over k+N}  p \cdot  \mu \biggr]
| p + n \mu\rangle. \label{knotop} \ee In this equation, $h_{\rho
+ \Lambda}$ is the conformal weight, and $M_{\Lambda}$ is the set
of weights corresponding to the irreducible representation with
highest weight $\Lambda$. This equation allows us to compute the
vev of the Wilson loop around a torus knot in ${\bf S}^3$ as
follows: first of all, one makes a Heegard splitting of
 ${\bf S}^3$ into two solid tori, as we explained before.
Then, one puts the torus knot on the
surface of one of the solid tori by acting with the knot operator
(\ref{knotop}) on the vacuum $| \rho\rangle$. Finally, one glues together
the tori by performing an
$S$-transformation. The normalized vev of the Wilson loop is then given by:
\be
\label{vevop}
\langle W_{ \Lambda}^{(n,m)}\rangle =
{\langle \rho | S W_{\Lambda}^{(n,m)}
| \rho\rangle \over \langle  \rho | S| \rho\rangle}.
\ee
One can show that \cite{knotops}
\be
W_{\Lambda}^{(1,0)}|\rho\rangle =|\rho + \Lambda \rangle.
\ee
On the other hand, the operator $W_{\Lambda}^{(1,0)}$ clearly creates a
trivial knot, or {\it unknot}, on the torus,
therefore the states $|\rho + \Lambda \rangle$
are obtained by doing the path integral over the solid torus with an
insertion of a Wilson loop around the noncontractible loop in the
representation $\Lambda$, as shown in \cite{cs}. We can now evaluate easily
the corresponding Chern-Simons invariant. Using the explicit expression in
(\ref{st}), we find:
\be
W_{R_{\Lambda}}({\rm unknot}) =
{\langle \rho| S W_{\Lambda}^{(1,0)} |\rho\rangle
\over \langle \rho| S  |\rho\rangle} =
{\sum_{w \in {\cal W}} \epsilon(w) {\rm e}^{-{2 \pi i \over k+N} \rho
\cdot w(\Lambda + \rho)} \over \sum_{w \in {\cal W}}
\epsilon(w) {\rm e}^{-{2 \pi i \over k+N} \rho
\cdot w( \rho)}}.
\ee
Using Weyl's denominator formula,
the vacuum expectation value can be written as
a character
\be
\label{qdimchar}
W_{R_{\Lambda}}({\rm unknot})=
{\rm ch}_{\Lambda}\Bigl[ -{2 \pi i \over k+N} \rho \Bigr].
\ee
Moreover, using (\ref{wdf}), we can finally write
\be
W_{R_{\Lambda}}({\rm unknot})=\prod_{\alpha >0} {\sin \Bigl( { \pi \over k+N}
\alpha\cdot (\Lambda + \rho) \Bigr) \over
\sin \Bigl( { \pi \over k+N}
\alpha\cdot \rho \Bigr)}.
\ee
Notice that, in the limit $k+N \rightarrow \infty$ ({\it i.e.} in the
semiclassical limit), this becomes the dimension of the representation
$R$. For this reason, the above quantity is called the {\it quantum
dimension} of $R$, denoted by ${\rm dim}_q R$. It can be explicitly written as
follows. Define the $q$-numbers:
\be
\label{qnumbers}
[x]=q^{x\over 2} -q^{-{x\over 2}},\,\,\,\ [x]_{\lambda} = \lambda^{1\over 2}
q^{x\over 2} -\lambda^{-{1\over 2}} q^{-{x\over 2}}.
\ee
If $R$ has a Young tableau with $c_R$ rows of lengths $l_i$,
$i=1, \cdots, c_R$,
then the quantum dimension can be explicitly written as:
\be
\label{expf}
{\rm dim}_q R = \prod_{1\le i < j \le c_R} {[l_i -l_j +j-i]
\over [j-i]} \prod_{i=1}^{c_R} { \prod_{v=-i+1}^{l_i -i} [v]_{\lambda}
\over \prod_{v=1}^{l_i} [v-i + c_R]}.
\ee
This gives the Chern-Simons invariant of the unknot in the representation $R$.

What about other torus knots? When
acting  with the knot operator (\ref{knotop}) on the vacuum,
we get the set of  weights
 $\rho + n \mu$, where $ \mu \in M_{ \Lambda}$. These
weights will have representatives in the Weyl alcove
${\cal F}_l$, which can be
obtained by a series of Weyl reflections. The set of
representatives in ${\cal F}_l$
 will be denoted by $ {\cal M} (n, \Lambda)$,
and it depends on the irreducible representation with
 highest weight $\Lambda$, and on the integer number
 $n$. Using the fact that $\rho + n \mu = w( \rho + \xi) $ for
some $w \in {\cal W}$, we conclude that the Chern-Simons
invariant of a torus
knot $(n,m)$ can be
written as:
\be
\label{vevch}
 {\rm e}^{2\pi i nm h_{ \rho +  \Lambda} }
\sum_{\rho+ \xi \in {\cal M}(n,  \Lambda)}
\exp \biggl[
-{i\pi m \over n(k+N)} \xi \cdot (\xi + 2 \rho)
\biggr] {\rm ch}_{ \xi}\Bigl[ -{ 2\pi i \over k+N} \rho\Bigr].
\ee
Notice that, since the representatives $\rho + \xi$ live in
${\cal F}_l$, the weights $\xi$ can
be considered as highest weights for a representation, hence (\ref{vevch})
makes sense. As an example of this procedure, one can compute the invariant
in the fundamental representation. By performing Weyl reflections, one can
show that $ {\cal M}(n, \lambda_1)$ is given by the
following weights \cite{homflytorus}:
\be
\label{fundweyl}
\rho + (n-i)\lambda_1 + \lambda_i,\,\,\,\,\,\,\,\, i=1, \cdots, N.
\ee
The computation of the characters is now straightforward (they are just the
quantum dimensions of the weights (\ref{fundweyl})), and one finally
obtains:
\be
 W_{\tableau{1}}^{(n,m)}
=t^{{1\over 2}}\lambda^{-{1\over 2}}
{(\lambda t^{-1})^{(m-1)(n-1)\over 2}\over t^n-1}
\sum_{p+i+1=n \atop  p, i \ge 0}(-1)^i t^{-mi+{1 \over
2} p(p+1) }{\prod _{j=-p}^i (\lambda -t^j) \over (i)! (p)!}
\ee
This is in fact the unnormalized
HOMFLY polynomial of an $(n,m)$ torus knot. If we divide by the
vev of the unknot, we find the expression for the HOMFLY
polynomial first obtained in \cite{jonesann}. For the trefoil one has for example:
\be
 W_{\tableau{1}} = {1 \over q^{1\over 2}
 -q^{-{1\over 2}} }(-2  \lambda^{1\over2} + 3  \lambda^{3\over2}
- \lambda^{5\over2}) + (q^{1\over 2} - q^{-{1\over 2}})(-\lambda^{1\over2} +
\lambda^{3\over2}).
\ee
With more effort one can obtain invariants of torus knots and
links in arbitrary representations \cite{homflytorus,lm,lmv}. For the trefoil in
representations with two boxes one finds:
\ben
W_{\tableau{2}}&=&
 {(\lambda-1)(\lambda q-1)  \over
 \lambda ( q^{{1\over 2}} -
 q^{-{1\over 2}}) ^2 \,( 1 + q)}
\Bigl((\lambda q^{-1})^2
( 1 - {\lambda}q^2 +q^3 \nonumber\\
& -&\lambda q^3+ q^4 -\lambda q^5
+ \lambda^2 q^5 +q^6 -\lambda q^6) \Bigr)\nonumber\\
W_{\tableau{1 1}}&=&
{(\lambda-1)(\lambda-q) \over \lambda  (  q^{{1\over 2}} -
 q^{-{1\over 2}}) ^2\,\,( 1 + q )  }
\Bigl( (\lambda q^{-2})^2 (1 -{\lambda} - \lambda q \nonumber\\
& + &{{{\lambda}}^2} q
+ q^2 +q^3 - {\lambda}q^3 -{{\lambda}}\,q^4 +q^6 ) \Bigr)
\label{trefoil}
\end{eqnarray}
For the Hopf link, one finds:
\begin{equation}
W_{(\tableau{1},\tableau{1})}=\biggl( {\lambda^{1\over2}-\lambda^{-
{1\over2}}
\over q^{1\over2} -q^{-{1\over 2}}}\biggr)^2-\lambda^{-1}(\lambda-1),
\label{hopf}
\end{equation}
which can be also easily obtained using the skein relations of the HOMFLY
polynomial (\ref{skeinrel}) together with  (\ref{homflyrel}).

\subsection{Framing dependence}
In the above discussion on the correlation functions of Wilson loops
we have missed an important ingredient. We mentioned that, in order to
define the partition function of Chern-Simons theory at the quantum level,
one has to specify a framing of the three-manifold. It turns out that the
evaluation of correlation functions like (\ref{vevknot}) also involves a
choice of framing of the knots, as Witten discovered in \cite{cs}. Since
this is important in the duality with topological strings, we will explain
it in some detail.

A good starting point to understand the framing is to take
Chern-Simons theory with gauge group $U(1)$. This is also useful
to understand $U(N)$ versus $SU(N)$ Chern-Simons theory, and to
get a concrete feeling of how to deal with correlation functions
like (\ref{vevknot}). The Abelian Chern-Simons theory turns out to
be extremely simple, since the cubic term in (\ref{csact}) drops
out, and we are left with a Gaussian theory \cite{polyakov}. The
different representations are labelled by integers, and in
particular the vevs of Wilson loop operators can be computed
exactly. In order to compute them, however, one has to choose a
framing for each of the knots ${\cal K}_i$. This arises as
follows: in evaluating the vev, contractions of the holonomies
corresponding to different ${\cal K}_i$ produce the following
integral:
\begin{equation}
\label{linking}
{\rm lk} ({\cal K}_i, {\cal K}_j)=
{1 \over 4 \pi} \oint_{{\cal K}_i}dx^{\mu} \oint_{{\cal K}_j} dy^{\nu}
\epsilon_{\mu \nu \rho} { (x-y)^{\rho} \over |x-y|^3}.
\end{equation}
This is in fact a topological invariant, {\it i.e.} it is invariant
under deformations of the knots ${\cal K}_i$, ${\cal K}_j$, and it is in fact
their linking number ${\rm lk}({\cal K}_i, {\cal K}_j)$.
On the other hand, contractions of the holonomies corresponding to the
same knot $\cal K$ involve the integral
\begin{equation}\label{cotor}
\phi ({\cal K})={1 \over 4 \pi} \oint_{\cal K}dx^{\mu} \oint_{\cal K} dy^{\nu}
\epsilon_{\mu \nu \rho} { (x-y)^{\rho} \over |x-y|^3}.
\end{equation}
This integral is well-defined and finite (see,
for example, \cite{guada}), and it is
called the cotorsion of $\cal K$. The problem is that the cotorsion
is not invariant under deformations of the knot. In order to
preserve topological invariance one has to choose another
definition of the composite operator $(\int_{\cal K}A)^2$ by means of a
framing. A framing of the knot consists of
choosing another knot ${\cal K}^f$ around $\cal K$,
specified by a normal vector
field $n$. The cotorsion $\phi({\cal K})$ becomes then
\begin{equation}
\label{regul}
\phi_f ({\cal K})={1 \over 4 \pi}
\oint_{\cal K}dx^{\mu} \oint_{{\cal K}^f} dy^{\nu}
\epsilon_{\mu \nu \rho} { (x-y)^{\rho} \over |x-y|^3} =
{\rm lk} ({\cal K}, {\cal K}^f).
\end{equation}
The correlation function that we obtain in this way is
a topological invariant (a linking number) but the
price that we have to pay is that our regularization depends on a set
of integers $p_i ={\rm lk} ({\cal K}_i, {\cal K}^f_i)$ (one for each knot).
The vev (\ref{vevknot}) in the Abelian case
can now be computed, after choosing the framings, as
follows:
\begin{equation}
\label{vevans}
\langle \prod_i \exp \bigl( n_i \int_{\gamma_i} A \bigr) \rangle =
\exp \biggl( { \pi i \over k} \sum_i n_i^2 p_i + {\pi i \over k}
\sum_{i \not= j} n_i n_j {\rm lk} ({\cal K}_i, {\cal K}_j) \biggr).
\end{equation}
This regularization is nothing but the `point-splitting' method
familiar in the context of QFT's.

Let us now consider Chern-Simons theory with gauge group $SU(N)$, and
suppose that you want to compute a correlation function like
(\ref{vevknot}). If you try to do it in perturbation theory, for example,
you will find very soon that self-contractions of the holonomies lead to
the same kind of ambiguities that we found in the Abelian case, {\it i.e.}
you will have to make a choice of framing for each knot ${\cal K}_i$. The
only difference is that the self contraction comes with a group factor ${\rm
Tr}_{R_i}(T_a T_a)$ for each knot ${\cal K}_i$,
where $T_a$ is a basis of the Lie algebra \cite{guada}. The precise result
can be better stated as the effect on the correlation function
(\ref{vevknot}) under a change of framing, and it says that,
under a change
of framing of ${\cal K}_i$ by $p_i$ units, the vev of the product of
Wilson loops changes as follows \cite{cs}:
\begin{equation}
\label{naframing}
W_{(R_1, \cdots, R_L)}  \rightarrow \exp \biggl[ 2\pi i \sum_i
p_i  h_{R_i} \biggr]W_{(R_1, \cdots, R_L)}
,
\end{equation}
In this equation, $h_R$ is the conformal weight of the WZW primary
field corresponding to the representation $R$. In
(\ref{confweight}) we labelled $R$ through $\alpha=\rho +
\Lambda$, where $\Lambda$ is the highest weight of $R$. In fact,
one can write (\ref{confweight}) as
\begin{equation}
\label{cweight}
h_R = {C_R\over 2(k+N)},
\end{equation}
where $C_R={\rm Tr}_{R}(T_a T_a)$
is the quadratic Casimir in the
representation $R$. For $SU(N)$,
one has
\begin{equation}
\label{explcasun}
C_R^{SU(N)} = N \ell + \kappa_R -{\ell^2 \over N},
\end{equation}
where $\ell$ is the total number of boxes in the tableau, and
\begin{equation}\label{kapar}
\kappa_R =\ell +  \sum_i \bigl( l_i^2 -2il_i  \bigr).
\end{equation}
We then see that the evaluation of vacuum
expectation values of Wilson loop operators in Chern-Simons theory depends
on a choice of framing for knots. It turns out that for
knots and links in ${\bf S}^3$, there is a {\it standard} or canonical
framing, defined by requiring that the self-linking number is zero. The
expressions listed in (\ref{trefoil}) and (\ref{hopf}) are all in the
standard framing, and the skein relations for the HOMFLY polynomial produce
invariants in the standard framing as well. Once the value of the invariant
is known in the standard framing, the value in any other framing specified
by nonzero integers $p_i$ can be easily obtained from (\ref{naframing}).

Let us now consider a $U(N)$ Chern-Simons theory. The $U(1)$
factor decouples from the $SU(N)$ theory, and all the vevs
factorize into an $U(1)$ and an $SU(N)$ piece. Representations of
$U(N)$ are also labelled by Young tableaux, and they decompose
into a representation of $SU(N)$ corresponding to that tableau,
and a representation of $U(1)$ with charge:
\begin{equation}\label{chargeuone}
n ={\ell \over {\sqrt N}},
\end{equation}
where $\ell$ is the number of boxes in the Young tableau. In order
to compute the vevs associated to the $U(1)$ of $U(N)$, one
has to take also into
account that the coupling constant $k$ is shifted as $k \rightarrow k+N$.
We then find that the vev of a product of $U(N)$ Wilson loops in
representations $R_i$ is given by:
\begin{equation}\label{factoriz}
W_{(R_1, \cdots, R_L)}^{U(N)} =
\exp \biggl( { \pi i \over N(k+ N) } \sum_i \ell_i^2 p_i +
{\pi i \over N(k +N)}
\sum_{i \not= j} \ell_i \ell_j {\rm lk} ({\mathcal K}_i, {\mathcal K}_j) \biggr)
W_{(R_1, \cdots, R_L)}^{SU(N)},
\end{equation}
where the $SU(N)$ vev is computed in the framing specified by $p_i$. Notice
that, in the case of knots, the $SU(N)$ and
$U(N)$ computations differ in a factor which only depends on the
choice of framing, while for links the answers also differ in a
topological piece involving the linking numbers. The change of framing for
vacuum expectation values in the $U(N)$ theory is again governed by
(\ref{naframing}) and (\ref{cweight}), but now the
quadratic Casimir is given by
\begin{equation}
\label{explcas}
C^{U(N)}_R = N \ell + \kappa_R,\end{equation}
Notice that the difference between the change of $SU(N)$ and
$U(N)$ vevs under the change of framing is consistent with (\ref{factoriz}).
In terms of the variables (\ref{varias})
we see that $U(N)$ vevs change, under the change of
framing, as
\begin{equation}
 \label{unframing}
W_{(R_1, \cdots, R_L)} \rightarrow  q^{{1 \over 2}\sum_i
 \kappa_{R_i} p_i } \lambda^{{1\over 2}
\sum_i \ell_i p_i} W_{(R_1, \cdots, R_L)}.
\end{equation}

\subsection{Generating functionals for Wilson loops}

As we will see, the relation between Chern-Simons theory and
string theory involves the vacuum expectation values for arbitrary
irreducible representations of $U(N)$, so it is convenient to have
a generating functional that encodes all the information about
them. We will for simplicity consider the case in which one has
just a single knot. We then have to find a suitable basis for the
Wilson loop operators. There are two natural basis for the
problem: the basis labelled by representations $R$, and the basis
labelled by conjugacy classes $C(\vec k)$ of the symmetric group.
Let $U$ be the holonomy of the gauge connection around the knot
${\cal K}$, and consider the operator $\Upsilon_{\vec k}(U)$
defined as in (\ref{ups}). The vevs of these operators give the
``$\vec k$-basis'' for the vacuum expectation values of the Wilson
loops:
\begin{equation}
W_{\vec k} = \langle \Upsilon_{\vec k}(U)\rangle =
\sum_{R} \chi_R (C(\vec k))W_R
\end{equation}
\noindent
where $\chi_R$ are characters of the permutation group
$S_{\ell}$ in the representation $R$, and we have used
Frobenius formula
\begin{equation}
{\rm Tr}_R(U)= \sum_{\vec k} {\chi_R (C(\vec k)) \over z_{\vec k}}
\Upsilon_{\vec k}(U), \label{frob}
\end{equation}
and $z_{\vec k}$ has been defined in (\ref{zk}). If $V$ is a
$U(M)$ matrix (a ``source'' term), one can define the following
operator, which was introduced in \cite{ov} and is known sometimes
as the Ooguri-Vafa operator:
\begin{equation}
Z(U,V)=\exp\Bigl[ \sum_{n=1}^\infty {1 \over n} {\rm Tr}\, U^n\,
{\rm Tr }\, V^n\Bigr].
\label{ovop}
\end{equation}
When expanded, this operator can be written in the $k$-basis as follows,
\begin{equation}
Z(U,V)=1 + \sum_{\vec k} {1 \over z_{\vec k}} \Upsilon_{\vec k}(U)
\Upsilon_{\vec k}(V).
\end{equation}
We see that $Z(U,V)$
includes all possible Wilson loop operators $\Upsilon_{\vec k}(U)$
associated to a knot ${\cal K}$. One can also use Frobenius formula
to show that
\begin{equation}
\label{ovrep}
Z(U,V)=\sum_{R} {\rm Tr}_R(U) {\rm Tr}_R (V),
\end{equation}
where the sum over representations starts with the trivial one.
In $Z(U,V)$ we assume that
$U$ is the holonomy of a dynamical gauge field and
that $V$ is a source. The vacuum expectation value $Z(V)=\langle Z(U,V)
\rangle$ has then information about the vevs of the Wilson loop operators,
and by taking its logarithm one can define the connected vacuum
expectation values $W^{(c)}_{\vec k}$:
\begin{equation}
F_{\rm CS}(V)=\log Z(V)= \sum_{\vec k} {1 \over z_{\vec k}!}
W^{(c)}_{\vec k} \Upsilon_{\vec k}(V) \label{convevs}
\end{equation}
One has, for example:
$$
W^{(c)}_{(2,0,\cdots)}=\langle ({\rm Tr} U)^2\rangle
-\langle {\rm Tr} U\rangle^2=W_{\tableau{2}} + W_{\tableau{1 1}}
-W_{\tableau{1}}^2.
$$
The free energy $F_{\rm CS}(V)$, which is a generating functional for
connected vevs $W^{(c)}_{\vec k}$, will be the relevant object for the
duality with topological strings.

\sectiono{Chern-Simons theory and large $N$ transitions}

\subsection{The $1/N$ expansion}
\begin{figure}
\leavevmode
\begin{center}
\epsfysize=4cm
\epsfbox{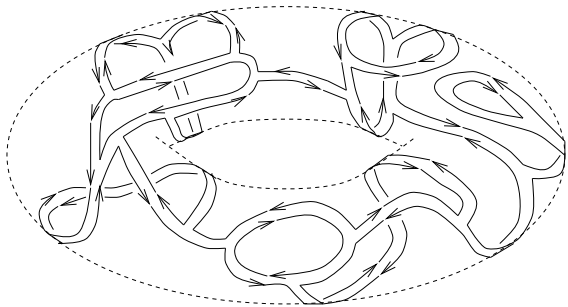}
\end{center}
\caption{This figure, taken from \cite{ovproof}, shows a fatgraph
with $h=9$ and $g=1$.}
\label{fatgraph}
\end{figure}

\begin{figure}
\leavevmode
\begin{center}
\epsfysize=4cm
\epsfbox{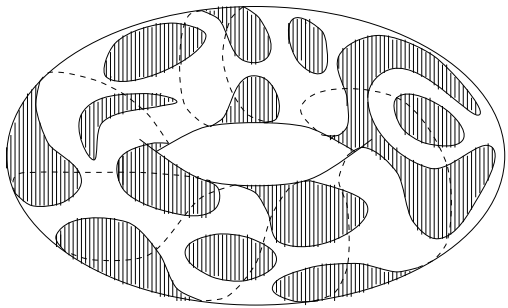}
\end{center}
\caption{The Riemann surface associated to the fatgraph of the
previous figure.}
\label{filling}
\end{figure}

As 't Hooft pointed out in \cite{thooft} (see \cite{coleman} for a nice
review),
given a theory with $U(N)$ or $SU(N)$ gauge symmetry one can always
perform a $1/N$ expansion of the free energy and the correlation functions.
To do that, one
writes the Feynman diagrams of the theory as ``fatgraphs'' or ribbon
graphs. The amplitude associated
to these ribbon graphs depends on the
coupling constant $x$ and on the rank of the gauge group (through its group
factor). Let us consider for example the expansion of the free energy. This
will involve connected vacuum bubbles with $V$ vertices, $E$ propagators
and $h$ loops of internal indices, and therefore will have a factor
\begin{equation}
x^{E-V}N^h=x^{2g-2+h}N^h=x^{2g-2}t^h,
\end{equation}
where $t=Nx$ is the so
called {\it 't Hooft parameter}. In writing this equation we
 regard the fatgraph as a Riemann surface with holes, {\it i.e.} each
internal loop represents the boundary of a hole, and we used Euler's
relation $E-V+h=2g-2$. In \figref{fatgraph} we show a fatgraph with $g=1$ and
$h=9$, and in \figref{filling} the Riemann surface that can be associated to
it.
 We can then write,
\begin{equation}
F^{\rm p}=\sum_{g=0}^{\infty} \sum_{h=1}^{\infty} F^{\rm p}_{g,h} x^{2g-2} t^h.
\label{openf}
\end{equation}
The superscript p means that this is the perturbative contribution to the
free energy. The full free energy may also have a
nonperturbative contribution. This is easily seen, in
the case of Chern-Simons theory, in (\ref{perts}): the
free energy has a perturbative contribution coming from the $S_{\ell}$, but
there is a nonperturbative contribution due to the one-loop prefactor (which also
depends on $N$, $x$) and involves one-loop determinants as well as
the precise normalization of the path integral.
In (\ref{openf}) we have written the diagrammatic series as an
expansion in $x$ around $x=0$, keeping $t$ fixed. Equivalently, we can
regard it as an
expansion in $1/N$ for fixed $t$, and then the $N$ dependence appears
as $N^{2g-2}$. The above expansion can be interpreted as the perturbative
expansion of an {\it open} string theory, where
$F^{\rm p}_{g,h}$ corresponds to
some amplitude on a Riemann surface of genus $g$ with $h$ holes like the
one depicted in \figref{filling}.
If we now introduce the function
\begin{equation}
F^{\rm p}_g (t) =\sum_{h=1}^{\infty} F^{\rm p}_{g,h} t^h,
\label{opencl}
\end{equation}
the total perturbative free energy can be written as
\begin{equation}
F^{\rm p}(x, t)=\sum_{g=0}^{\infty} x^{2g-2} F^{\rm p}_g(t),
\label{closedf}
\end{equation}
which looks like a {\it closed} string expansion where $t$ is some
modulus of the theory. Notice that in writing (\ref{opencl}) we have
grouped together all open Riemann surfaces with the same bulk topology but with
different number of holes, so by ``summing over all holes'' we
``fill up the holes'' to produce a closed Riemann surface. This leads to 't
Hooft's idea \cite{thooft} that, given a gauge theory, one
should be able to find a string
theory interpretation in the way we have described, namely, the fatgraph
expansion of the free energy is resummed to give a function of the 't
Hooft parameter $F_g(t)$ at every genus that is then interpreted as a
closed string amplitude.

We can now ask what is the interpretation of the vacuum expectation values
of Wilson loop operators in this context. Using standard large $N$
techniques (as reviewed for example in \cite{coleman}), it is easy to see that the vevs that have a well-defined
behavior in the $1/N$ expansion are the connected vevs $W_{\vec k}^{(c)}$
introduced in (\ref{convevs}). One finds that these vevs admit an expansion
of the form,
\begin{equation}
  W_{\vec k}^{(c)}= \sum_{g=0}^{\infty} W_{g,
 \vec k} (t)  x^{2g-2 + |\vec k|}.
\label{opencon}
\end{equation}
This can be regarded
as an open string expansion, where $W_{g,\vec k}(t)$ are interpreted as
amplitudes in an open string theory at genus $g$ and with $h=| \vec k|$
holes. The vector $\vec k$ specifies the winding numbers of the holes around a
one-cycle in the target space of the theory, according to the rule we gave
in subsection 2.3. We could say that the Wilson loop
``creates'' a one-cycle in the target
space where the boundaries of Riemann surfaces can end, and the generating
functional for connected vevs (\ref{convevs}) is interpreted as the total
free energy of an open string, as in (\ref{totalfreeopen}). These open
strings shouldn't be confused with the ones that we associated to the
expansion (\ref{openf}). The open strings underlying (\ref{opencon}) should
be regarded as an open string sector in the closed string theory associated
to the resummed expansion (\ref{closedf}).

This is then the program to interpret gauge theories with $U(N)$ or $SU(N)$
symmetry in terms of a string theory. So far this program has been
led to completion in just a few examples. A first example is a class of
gauge theories in
zero dimensions, the matrix models of Kontsevich, which are equivalent to
topological minimal matter in two dimensions coupled to gravity \cite{k},
{\it i.e.} to topological strings in $d<1$ dimensions. Another
example is Yang-Mills theory in two dimensions, which also has a string
theory description \cite{gt,cmr}. Finally, ${\cal N}=4$
supersymmetric Yang-Mills
theory is equivalent to type IIB string theory on ${\bf S}^5 \times
{\rm AdS}_5$
\cite{malda}. The last example shows very clearly that the target of the
string theory is not necessarily the spacetime where the gauge
theory lives, and that
the string description may need ``extra" dimensions. The question
we want to address now is the following: is there a
string description of Chern-Simons theory? As we will see,
at least for Chern-Simons
on the three-sphere, the answer is yes. The resulting description provides
a very nice realization of 't Hooft ideas, and as
we will show, leads to new
insights on knot and link invariants\footnote{Other attempts to find a
string theory interpretation of Chern-Simons theory can be found in
\cite{peri,douglas}.}.

\subsection{Chern-Simons theory as an open string theory}

In order to give a string theory interpretation of Chern-Simons theory
on ${\bf S}^3$, a
good starting point is to give an open string interpretation to the $1/N$
expansion of the free energy (\ref{openf}). This was done by
Witten in \cite{csts}, and we will summarize here the main points of the
argument.

First of all, we have to recall that open bosonic strings have
 a spacetime description
in terms of the cubic open string field theory introduced in
\cite{sft}. The action of this theory is given by
\be
S={1 \over g_s} \int \biggl( {1 \over 2} \Psi \star Q_{\rm BRST} \Psi + {1 \over 3} \Psi \star \Psi
\star \Psi \biggr).
\label{cubicsft}
\ee
In this equation, $\Psi$ is the string field, $\star$ is the associative,
noncommutative product obtained by gluing strings, and the integration is
a map $\int : \Psi \rightarrow {\bf R}$ that involves the gluing of the two
halves of the string field (more details can be found in \cite{sft}). If we
add Chan-Paton factors, the string field is promoted to a $U(N)$ matrix of
string fields, and the integration includes ${\rm Tr}$. This action has all
the information about the spacetime dynamics of open bosonic strings, with
or without D-branes. In
particular, one can derive the Born-Infeld action describing the dynamics
of D-branes from this cubic string field theory (see for
example \cite{taylor}).

Consider now a three-manifold $M$. The total space of its
cotangent bundle $T^*M$ is a noncompact Calabi-Yau manifold.
Moreover, it is easy to see that $M$ is a Lagrangian submanifold
in $T^*M$. We can then consider a system of $N$ topological
D-branes wrapping $M$, thus providing Dirichlet boundary
conditions for the open strings. We want to obtain a spacetime
action describing the dynamics of these topological D-branes. To
do this, we can exploit again the analogy between open topological
strings and the open bosonic string that we used to define the
coupling of topological sigma models to gravity ({\it i.e.}, that
both have a nilpotent BRST operator and an energy-momentum tensor
that is $Q_{\rm BRST}$-exact). Using the fact that both theories
have a similar structure, one can argue \cite{csts} that the
dynamics of topological D-branes in $T^*M$ is governed as well by
(\ref{cubicsft}). However, one has to work out what is exactly the
string field, the $\star$ algebra and so on in the context of
topological open strings. It turns out that the string field is
simply a $U(N)$ gauge connection $A$ on $M$, the integration of
string functionals becomes ordinary integration of forms on $M$,
and the star product becomes the usual wedge product of forms. We
then have the following dictionary: \be
\begin{array}{ccc}
 \Psi \rightarrow A, & \,  & Q_{\rm BRST}\rightarrow d\\
\,  & \,  &\, \\
\star \rightarrow \wedge, & \,  & \int \rightarrow \int_M.
\end{array}
\ee
The resulting action (\ref{cubicsft}) is then the usual
Chern-Simons action, and we have the following relation
between the string coupling constant
and the Chern-Simons coupling
\be
g_s ={2 \pi \over k+N},
\label{oscc}
\ee
after accounting for the usual shift $k \rightarrow k +N$.
Notice that, in the open bosonic string, the string field
involves an infinite tower of string excitations. For the open topological
string, the topological character of the model implies that all excitations
are $Q$-exact (and therefore decouple), except for the lowest
lying one, which is a $U(N)$ gauge connection. In other words, the usual
reduction to a finite number of degrees of freedom that occurs in
topological theories downsizes the string field to a single excitation.

The topological open string theory that we are obtaining has some
important differences with the one that we described in section 2.
As Witten pointed out in \cite{csts}, there are no honest
worldsheet instantons in this geometry! To be precise, worldsheet
instantons whose boundaries lie in $M$ must have zero area, and one
would then conclude that the only contributions come from constant
maps. A detailed analysis shows however that there are nontrivial
worldsheet instantons contributing to the path integral, but they
are degenerate and belong to the boundary of the moduli space of
holomorphic maps. These degenerate instantons look just like
fatgraphs, and in fact they correspond to the Feynman diagrams of the
$1/N$ expansion of Chern-Simons theory! In particular, to
characterize topologically these degenerate instantons we just
need their genus $g$ and number of holes $h$, which are of course
the same ones of the associated fatgraph. There are no winding
numbers to specify.

The outcome of this discussion is that, for topological open strings on
noncompact Calabi-Yau manifolds
of the form $T^*M$, the dynamics is governed by the usual Chern-Simons
action on $M$. In particular, the coefficient
$F^{\rm p}_{g,h}$ in (\ref{openf}) can be
interpreted as the free energy of an open string of genus $g$ and $h$ holes
propagating on $T^*M$ and with Lagrangian boundary conditions specified by
$M$.

\begin{figure}
\leavevmode
\begin{center}
\epsfysize=3.5cm
\epsfbox{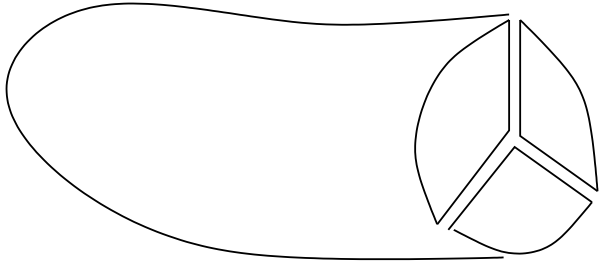}
\end{center}
\caption{This figure shows a partially degenerated worldsheet instanton of genus $g=0$ and
with $h=3$ ending on an unknot. The instanton is made out of a honest
holomorphic disk and the degenerate piece, which is a fatgraph.}
\label{first}
\end{figure}

This result can be extended \cite{csts}, and the more general picture will
be extremely useful later on. Consider a Calabi-Yau
manifold $X$ together with some Lagrangian submanifolds $M_i \subset X$,
with $N_i$ D-branes wrapped over $M_i$.
In this case the topological open strings will have contributions
from degenerate holomorphic curves, which are captured
by Chern-Simons theories in the way we explained for $T^*M$,
as well as some honest holomorphic curves. As shown in \cite{csts},
these honest holomorphic curves are open Riemann surfaces whose boundaries
are embedded knots inside the three-manifolds $M_i$
and give rise to Wilson loops. Each holomorphic curve with area
$A$ ending on $M_i$ over the knot ${\cal K}_i$
will contribute $e^{-A} \prod_i {\rm Tr}U_{{\cal K}_i}$
to the free energy, where $U_{{\cal K}_i}$ denotes the holonomy of the
Chern-Simons $U(N_i)$ gauge connection $A_i$ around the knot ${\cal K}_i$.
We can then take into account the contributions of all curves
by including the corresponding Chern-Simons theories
$S_{\rm CS}(A_i)$, which account for the degenerate curves,
coupled in an appropriate way to the honest holomorphic
curves. The spacetime action will then have the form
\be
\label{impo}
S(A_i)= S_{\rm CS}(A_i)+F_{\rm ndg}(U_{{\cal K}_i})
\ee
where
\be
F_{\rm ndg}=\sum_{\rm instantons} e^{-A} \prod_i {\rm Tr}U_{{\cal K}_i}
\label{nondeg}
\ee
denotes the contribution
of the non-degenerate holomorphic curves, and it is a sum over honest open
worldsheet instantons. Notice
that all the Chern-Simons theories $S_{\rm CS}(A_i)$
have the same coupling constant, equal to the string coupling constant.
More precisely,
\be
\label{sameness}
{2\pi \over k_i+N_i}=g_s.
\ee
In the action (\ref{impo}), the honest holomorphic curves are
put ``by hand'' in $F_{\rm
ndg}$, and in principle one has to solve a nontrivial enumerative problem
to find them. Once they are included in the action, the path integral over
the Chern-Simons connections will join degenerate instantons
to these honest worldsheet
instantons: if we have a nondegenerate
 worldsheet instanton ending on a knot ${\cal K}$,
it will give rise to a Wilson loop operator in (\ref{nondeg}), and the
evaluation of the vacuum expectation value will generate,
in the $1/N$ expansion, all possible
fatgraphs $\Gamma$ joined to the knot ${\cal K}$, as it is well-known in
Chern-Simons perturbation theory in the presence of Wilson loops (see for example
\cite{tenyears}). These fatgraphs are interpreted
as degenerate instantons. Therefore, the path
integral with the action (\ref{impo})
will be a sum of contributions coming from
partial degenerations of Riemann surfaces, in which a surface
$\Sigma_{g,h}$ degenerates to another surface $\Sigma_{g',h'}$ whose
boundary ends on a knot ${\cal K}$, together with a fatgraph whose external
legs end in ${\cal K}$ as well. An example of this situation is depicted in
\figref{first}, where a disc ends on an unknot, and the fatgraph generated
by Chern-Simons perturbation theory gives in the end a Riemann surface of
$g=0$ and $h=3$. This more complicated scenario
was explored in \cite{avg2,dfg,dfg2,amv}, and we will provide examples of
(\ref{impo}) in section 6.

\subsection{The conifold transition}

We have learned that Chern-Simons theory on ${\bf S}^3$ is a
topological open string theory on $T^*{\bf S}^3$. Notice that the target of the
string theory is different from (and has higher dimensionality than)
the spacetime of the gauge theory, as in the string
description of ${\cal N}=4$ Yang-Mills theory. The next step is to see
if there is a {\it closed} string theory leading to the resummation
(\ref{closedf}). As shown by Gopakumar and Vafa in an important
paper \cite{gvgeom}, the answer is yes.

One way to motivate their result is as follows: since the holes of the Riemann surfaces are due to the
presence of D-branes, ``filling the holes'' to get the closed strings means
getting rid of the D-branes. But this is precisely what happens in
the AdS/CFT correspondence \cite{malda},
where type IIB theory in flat space in the presence of D-branes is
conjectured to be equivalent to type IIB theory in ${\rm AdS}_5 \times {\bf
S}^5$ with no D-branes. The reason for that is that, at large $N$, the
presence of the D-branes can be traded by a deformation of the background
geometry, and the radius of the ${\bf S}^5$ is related
to the number of D-branes. In other words,
we can make the branes disappear if we change the
background geometry at the same time. As Gopakumar and Vafa have pointed
out, large $N$ dualities relating open and closed strings should involve
transitions in the geometry. This reasoning suggests to look for a
transition involving the background
$T^* {\bf S}^3$. It turns out that such a transition is well-known in the
physical and the mathematical literature, and it is called the conifold
transition (see for example \cite{co}). Let us explain this in detail.

Although we have regarded $T^*{\bf S}^3$ as the total space of the
cotangent space bundle of the three-sphere, this background can be
also regarded as the deformed conifold geometry, which is usually described by
the algebraic equation
\be
\sum_{\mu=1}^4 \eta_{\mu}^2 =a.
\label{defconifold}
\ee
To see this equivalence, let us
write $\eta_{\mu}= x_{\mu} + i p_{\mu}$, where $x_{\mu}$,
$p_{\mu}$ are real coordinates.
We find the two equations
\ben
\sum_{\mu=1}^4 ( x_{\mu}^2 -p_{\mu}^2) & =& a, \nonumber\\
\sum_{\mu=1}^4 x_{\mu} p_{\mu}&=&0.
\een
The first equation indicates that the
locus $p_{\mu}=0$, $\mu=1, \cdots, 4$, describes a
sphere ${\bf S}^3$ of radius $R^2=a$, and the second equation shows that
the $p_{\mu}$ are coordinates for the cotangent space. Therefore,
(\ref{defconifold}) is nothing but $T^*{\bf S}^3$.

It is useful to rewrite the deformed conifold in yet
another way. Introduce the following complex coordinates:
\be
\begin{array}{ccc}
\label{newcoords}
x=\eta_1 + i \eta_2, & & v=i(\eta_3 -i \eta_4), \\
u=i(\eta_3 + i \eta_4), & & y= \eta_1 -i \eta_2.
\end{array}
\ee
The deformed conifold can be now written as
\be
xy=uv+a.
\label{defalt}
\ee
Notice that in this parameterization the geometry has a ${\bf T}^2$ fibration
\be
x,y,u,v \rightarrow xe^{i\theta_a},ye^{-i\theta_a},ue^{i\theta_b},ve^{-i\theta_b}
\label{torusact}
\ee
where the $\theta_a$ and $\theta_b$ actions above can be taken to generate
the $(1,0)$ and $(0,1)$ cycles of the ${\bf T}^2$. The ${\bf T}^2$
fiber can degenerate to ${\bf S}^1$ by collapsing
one of its one-cycles. In the equation above, for example,
the $U(1)_a$ action fixes $x=0=y$ and therefore fails to
generate a circle there.
In the total space, the locus where this happens, i.e. the $x=0=y$
subspace of $X$, is a cylinder $uv=-a$ . Similarly, the locus where the
other circle collapses, $u=0=v$, gives another cylinder $xy=a$.
Therefore, we can regard the whole
geometry as a ${\bf T}^2 \times\IR$ fibration over $\IR^3$: if we define
$z=uv$, the $\IR^3$ of the base is given by ${\rm Re}(z)$ and the axes
of the two cylinders. The fiber is given by the circles of the two
cylinders, and by ${\rm Im}(z)$. It is very useful to represent the above geometry by
depicting the singular loci of the torus action in the base $\IR^3$.
\begin{figure}
\leavevmode
\begin{center}
\epsfysize=6cm
\epsfbox{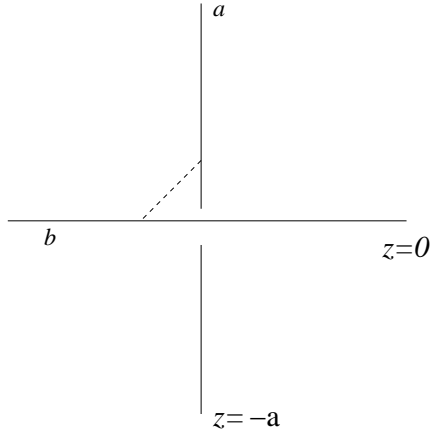}
\end{center}
\caption{This figure represents $T^*{\bf S}^3$, regarded as a ${\bf T}^2
\times \IR$ fibration of $\IR^3$. Two of the directions represent the axes
of the two cylinders, and the third direction represents the real axis of
the $z$-plane.}
\label{deformedf}
\end{figure}
The loci where the cycles of the torus collapse, which are cylinders,
project to lines in the base space. Notice that ${\bf S}^3$ can be regarded as a
torus fibration over an interval, with singular loci at the endpoints.
In \figref{deformedf}, the three-sphere of the deformed conifold geometry
is represented by a dashed line in the $z$-plane between $z=0$ and $z=-a$, together
with the $\theta_a$ and the $\theta_b$ circles that degenerate over the
endpoints.

\begin{figure}
\leavevmode
\begin{center}
\epsfysize=7cm
\epsfbox{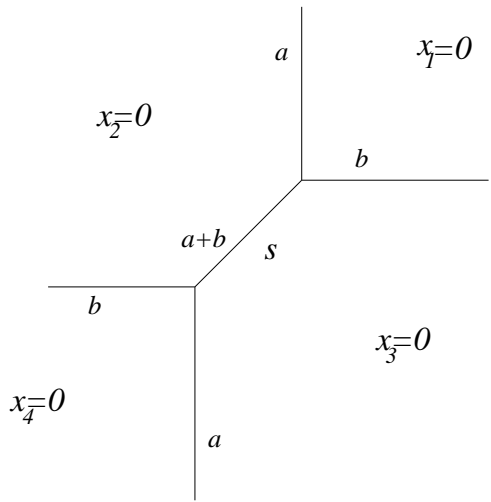}
\end{center}
\caption{This figure represents the resolved conifold ${\cal O}(-1) \oplus
{\cal O}(-1) \rightarrow \IP^1$ and the fixed point loci of the ${\bf T}^2$
action.}\label{resolvedf}
\end{figure}

The conifold singularity appears when $a=0$ and the three-sphere collapses.
This is described by the equation:
\be
\label{conifold}
xy=uv.
\ee
In algebraic geometry, singularities can be avoided in
two ways, in general. The
first way is to deform the complex geometry. This leads in our case to
the deformed conifold (\ref{defconifold}). The other way is to resolve
the singularity, for example by performing a blow up, and this leads to the
resolved conifold geometry (see for example \cite{co}). The resolution of
the geometry can be explained as follows. When $a=0$, (\ref{defalt}) says
that $xy=uv$. We can solve (\ref{conifold}) by setting
\be
x=\lambda v, \,\,\,\,\,\, u= \lambda y
\label{rescon}
\ee
where $\lambda$ is regarded as
an inhomogeneous coordinate in $\IP^1$. The space described by the complex
coordinates $x,y,\lambda,u,v$ together with the relations (\ref{rescon}) is the
resolved conifold, and it turns out to be
the bundle ${\cal O}(-1) \oplus {\cal O}(-1)
\rightarrow \IP^1$, as one can see from (\ref{rescon}) \cite{co}. To make
contact with the toric description
given in (\ref{toricres}), we put $x=x_1 x_2$, $y=x_3 x_4$, $u=x_1 x_3$ and
$v=x_2 x_4$. We then see that $\lambda=x_1/x_4$ is the inhomogeneous coordinate
for the $\IP^1$ described in (\ref{toricres}) by $|x_1|^2 + |x_4|^2=s$. It
is instructive to represent the resolved conifold by solving the constraint
(\ref{toricres}) in the first octant of $\IR^3$, and depicting the fixed
point locus of the isometries above. In terms of the coordinates $x_1,
\cdots, x_4$, the ${\bf T}^2$ action (\ref{torusact}) is given by
\be
x_1, x_2, x_3, x_4 \rightarrow {\rm e}^{i(\theta_a +\theta_b)} x_1,
{\rm e}^{-i\theta_a } x_2,{\rm e}^{-i\theta_b } x_3, x_4,
\ee
and the fixed loci are depicted in \figref{resolvedf}.
In the conifold transition, the three-sphere of the deformed conifold
shrinks to zero size as $a$ goes to zero, and then a two-sphere of size $s$
blows up giving the resolved conifold.

We know that Chern-Simons theory is an open topological string on the
deformed conifold geometry with $N$ topological D-branes wrapping the
three-sphere. The conjecture of Gopakumar and Vafa is that at large $N$ the D-branes
induce a conifold transition in the background geometry, so that we end up with the resolved
conifold and no D-branes. But in the absence of D-branes that enforce boundary
conditions we just have a theory of closed
topological strings. Therefore, {\it Chern-Simons theory on ${\bf S}^3$
is equivalent to closed topological string theory on the resolved conifold.}

This conjecture has been proved by embedding the
duality in type II superstring
theory \cite{vafa} and lifting it to
M-theory \cite{acharya,atmv}, and more recently
a worldsheet derivation
has been presented in \cite{ovproof}. In the remaining of this section, we will give
evidence for the conjecture at the level of the free energy.

\subsection{First test of the duality: the free energy}

A nontrivial test of the duality advocated by Gopakumar and Vafa is to
verify that the free energy of $U(N)$ Chern-Simons theory on the sphere agrees
with the free energy of closed topological strings on the resolved
conifold. The partition function of CS with gauge group $U(N)$ on the
sphere is a slight modification of (\ref{css3}):
\be
Z={1 \over (k+N)^{N/2}} \prod_{\alpha>0}
2 \sin \Bigl( {\pi (\alpha \cdot \rho) \over k+N} \Bigr).
\label{css3conc}
\ee
and differs from it in an overall factor $N^{1/2}/(k+N)^{1/2}$ which is the
partition function for the $U(1)$ factor (recall that $U(N)=U(1) \otimes
SU(N)/{\bf Z}_N$). Using the explicit description of the positive roots of
$SU(N)$, one gets
\be
F=\log Z= -{N \over 2} \log (k+N) +
\sum_{j=1}^{N-1} (N-j)\log \biggl[ 2 \sin { \pi j \over k+N} \biggr].
\label{csfs3}
\ee
We can now write the $\sin$ as
\be
\sin \pi z = \pi z \prod_{n=1}^{\infty}\biggl( 1 -{ z^2 \over n^2} \biggr),
\label{sinpro}
\ee
and we find that the free
energy is the sum of two pieces. One of them is the
nonperturbative piece:
\be
F^{\rm np}=-{N^2 \over 2}\log (k+N) + {1 \over 2}N(N-1) \log 2\pi +
 \sum_{j=1}^{N-1} (N-j) \log j,
\label{np}
\ee
and the other piece is the perturbative one:
\be
F^{\rm p}
=\sum_{j=1}^{N-1} (N-j) \sum_{n=1}^{\infty} \log \biggl[ 1 - {j^2  g_s^2
\over 4 \pi^2 n^2}\biggr],
\label{compli}
\ee
where $g_s$ corresponds to the open string coupling constant and it is
given by (\ref{oscc}).
To see that (\ref{np}) corresponds to the nonperturbative piece of the free
energy, we notice that the volume of $U(N)$ can be written as (see for
example \cite{ovproof}):
\be
{\rm vol}(U(N))={ (2\pi)^{ {1 \over 2}N(N+1)} \over G_2(N+1)}
\label{volun}
\ee
where $G_2(z)$ is the Barnes function, defined by
\be
G_2 (z+1)=\Gamma (z) G_2(z), \,\,\,\,\, G_2(1)=1.
\ee
It is now easy to see that
\be
F^{\rm np}=\log \, { (2 \pi g_s)^{{1 \over 2}N^2}
\over {\rm vol}(U(N))}
\ee
so it is given by the log of (\ref{asympt}), where $A^{(c)}$ is in this case
the trivial flat connection. Therefore, $F^{\rm np}$ is the log of the
prefactor associated to the normalization of
the path integral, which is not captured by Feynman diagrams.

Let us work out the perturbative piece (\ref{compli}).
By expanding the $\log$, using that $\sum_{n=1}^{\infty} n^{-2k}=\zeta
(2k)$, and the formula
\be
\sum_{j=1}^{N-1} j^{2k} =-{N^{2k}\over 2}+ \sum_{l=0}^{k}{2k +1
\choose 2l} {B_{2l} \over 2k+1} N^{2k+1-2l}
\ee
we find that (\ref{compli}) can be written as
\be
F^{\rm p} = \sum_{g=0}^{\infty} \sum_{h=2}^{\infty} F^{\rm p}_{g,h}
g_s^{2g-2+h} N^h,
\ee
where $F^{\rm p}_{g,h}$ is given by:
\ben
F^{\rm p}_{0,h}&=& -{2 \zeta(h-2)\over (2 \pi)^{h-2}(h-2)h(h-1)}, \nonumber\\
F^{\rm p}_{1,h}&=&{1 \over 6}{\zeta(h) \over (2 \pi)^{h} h  },   \nonumber\\
F^{\rm p}_{g,h}&=& 2 {\zeta (2g-2+h) \over (2 \pi)^{2g-2+h}}
{2g-3+h \choose h}{ B_{2g} \over 2g (2g-2)},\,\,\,\, g\ge 2.
\label{fghs3}
\een
This gives the contribution of connected diagrams with
two loops and beyond to the free energy of Chern-Simons on the sphere,
so we can write
\be
\sum_{\ell=1}^{\infty} S_{\ell}(N) x^l =\sum_{g=0}^{\infty}
\sum_{h=2}^{\infty}(-1)^{g-1+h/2}F_{g,h}x^{2g-2+h} N^h,
\ee
where $x$ is given by (\ref{coupling}), and we have explicitly
indicated the dependence of $S_{\ell}$ on $N$. Notice that the only
nonzero $F_{g,h}$ have $h$ even.
One can check that the $F_{g,h}$ that
we have obtained in (\ref{fghs3}) are in agreement with known results of
perturbative Chern-Simons theory on the sphere (see for example \cite{ww,csp}).
The nonperturbative piece also admits an expansion that
can be easily worked out from the asymptotics of the Barnes function
\cite{peri,ovproof}. One finds:
\be
F^{\rm np}= {N^2 \over 2} \Bigl( \log (Ng_s) -{3 \over 2} \Bigr)
-{1 \over 12}\log N + \zeta' (-1) + \sum_{g=2}^{\infty} {B_{2g} \over
2g (2g-2)} N^{2-2g}.
\ee

So far, what we have uncovered is the open string expansion of Chern-Simons
theory, which is (order by order in $x$) determined by the perturbative
expansion. In order to find a closed string interpretation, we have to sum
over the holes, as in (\ref{opencl}). Define the `t Hooft parameter $t$ as:
\be
t=ig_s N = xN,
\label{thooft}
\ee
then
\be
F^{\rm p}_g(t) = \sum_{h=1}^{\infty}F_{g,h}^{\rm p} (-it)^h.
\ee
We will now focus on $g\ge 2$.
To perform the sum explicitly, we write again the $\zeta$ function
as $\zeta(2g-2+ 2p)=\sum_{n=1}^{\infty} n^{2-2g-2p}$, and use the
binomial series,
\be
{1 \over (1-z)^q}=\sum_{n=0}^{\infty} {q+n-1 \choose n} z^n
\ee
to obtain:
\be
F^{\rm p}_g(t)= {(-1)^g  |B_{2g} B_{2g-2}| \over
2g (2g-2)(2g-2)!} +  {B_{2g} \over 2g (2g-2)} \sum_{n \in {\bf Z}} \,'
{1 \over (-it+ 2\pi n)^{2g-2}},
\label{quasi}
\ee
where $'$ means that we omit $n=0$. Now we notice that, if we write
\be
F^{\rm np}=\sum_{g=0}^{\infty} F^{\rm np}_g(t) g_s^{2g-2}
\ee
then for, $g\ge 2$,
$$F^{\rm np}_g(t)={ B_{2g}\over 2g(2g-2)}(-it)^{2-2g},
$$ which
is
precisely the $n=0$ term missing in (\ref{quasi}). We then define:
\be
F_g(t)=F_g^{\rm p}(t) + F_g^{\rm np}(t).
\ee
Finally, since
\be
\sum_{n \in {\bf Z}}{1 \over n+ z}={2 \pi i \over 1-{\rm e}^{-2\pi i z}},
\ee
by taking derivatives w.r.t. $z$ we can write
\be
F_g(t)={(-1)^g  |B_{2g} B_{2g-2}| \over
2g (2g-2)(2g-2)!} + {|B_{2g}| \over
2g (2g-2)!}{\rm Li}_{3-2g}({\rm e}^{-t}),
\label{fin}
\ee
again for $g\ge 2$. If we now compare (\ref{fin}) to (\ref{multibuble}),
we see that
it has {\it precisely} the structure of the free energy of a closed
topological string, with $n_1^0=1$, and the rest of the Gopakumar-Vafa
invariants being zero. Also, from the first term, which gives the
contribution of the constant maps, we find that $\chi(X)=2$. In fact,
(\ref{fin}) is the $F_g$ amplitude of the resolved conifold. One can also work out the
expressions for $F_0 (t)$ and $F_1 (t)$ and find agreement with the corresponding
results for the resolved
conifold \cite{gvgeom}. This
is a remarkable check of the conjecture.

\sectiono{Wilson loops and large $N$ transitions}

\subsection{Incorporating Wilson loops}

How do we incorporate Wilson loops in the large $N$ duality for Chern-Simons
theory? As we discussed in the previous section,
once one has a closed string description of the $1/N$ expansion,
Wilson loops are related to the open string sector in the closed string
geometry. Since the string description involves topological strings, it is
natural to assume that Wilson loops are going to be described by open
topological strings in the resolved conifold, and this means that we need a
Lagrangian submanifold specifying boundary conditions.

These issues were addressed in an important paper by Ooguri and
Vafa \cite{ov}. In order to give boundary conditions for the open
strings in the resolved conifold, Ooguri and Vafa constructed a
Lagrangian submanifold ${\widehat {\cal C}}_{\cal K}$ in $T^*{\bf
S}^3$ for any knot ${\cal K}$ in ${\bf S}^3$. This Lagrangian is
rather canonical, and it is called the conormal bundle of ${\cal
K}$. The details are as follows: suppose that the knot is
parameterized by a curve $q(s)$, where $s\in [0, 2\pi)$, for
example. The conormal bundle of ${\cal K}$ is then the space \be
{\widehat {\cal C}}_{\cal K}= \Bigl\{ (q(s), p) \in T^*{\bf S}^3 |
\sum_i p_i \dot q_i=0, \,\, 0 \le s< 2\pi \Bigr\} \ee where $p_i$
are coordinates for the cotangent bundle, and $\dot q_i$ denote
the derivatives w.r.t. $s$. This space is an $\IR^2$-fibration of
the knot itself, where the fiber on the point $q(s)$ is given by
the two-dimensional subspace of $T_q^* {\bf S}^3$ of planes
orthogonal to $\dot q(s)$. $ {\widehat {\cal C}}_{\cal K}$ has in
fact the topology of ${\bf S}^1 \times \IR^2$, and intersects
${\bf S}^3$ along the knot ${\cal K}$.

One can now consider, together with the $N$ branes wrapping ${\bf S}^3$, a
set of $M$ probe branes wrapping ${\widehat {\cal C}}_{\cal
K}$, and study the effective theory that one obtains in this way. On the
$N$ branes wrapping ${\bf S}^3$ we have $U(N)$
Chern-Simons theory. But the strings stretched between the $N$ branes and
the $M$ branes give an extra state in topological
string field theory, which turns out to be
a massless complex scalar field $\phi$ in the bifundamental
representation $(N, \overline M)$, and living in the intersection of the
two branes, ${\cal K}$. If $A$ denotes the $U(N)$ gauge connection on ${\bf
S}^3$, and $\widetilde A$ denotes the $U(M)$ gauge connection on
${\widehat {\cal C}}_{\cal
K}$, the action for the scalar is given by
\be
  \oint_{\cal K}\;
{\rm Tr} \,  \bar{\phi}\,D \phi, \ee where $D=d+ A  - {\widetilde
A}$. Here we regard $\widetilde A$ as a source. We can now proceed
to integrate out $\phi$ \cite{ov}. This is just a one loop
computation giving \be \exp \Bigl[ -\log \, {\rm det}\, D
\Bigr]=\exp \Bigl[-{\rm Tr} \log \Bigl( U^{-{1\over2}} \otimes
V^{1 \over2} - U^{1\over 2} \otimes V^{-{1\over2}}\Bigr)\Bigr].
\ee In this equation, $U$, $V$ are the holonomies of $A$,
$\widetilde A$ around the knot ${\cal K}$. To obtain this
equation, we have diagonalized $A$, $\tilde A$ and taken into
account that \be \log \, {\rm det}\Bigl[ {d \over ds} + i
\theta\Bigr]=\sum_{n=-\infty}^{\infty} \log(n+\theta) =\log \sin
(\pi \theta) + {\rm const.}, \ee where use has been made of
(\ref{sinpro}). In this way we obtain the effective action for the
$A$ field \be S_{\rm CS}(A) + \sum_{n=1}^{\infty}{1 \over n} {\rm
Tr}U^n {\rm Tr} V^{-n} \label{defact} \ee where $S_{\rm CS}(A)$ is
the Chern-Simons action for $A$ associated to the $N$ branes on
the three-sphere \footnote{In the above equation we have factored
out a contribution involving the $U(1)$ pieces of $U(N)$, $U(M)$.
These can be reabsorbed in a change of framing.}. Therefore, in
the presence of the probe branes, the action gets deformed by the
Ooguri-Vafa operator that we introduced in (\ref{ovop}). Since we
are regarding the $M$ branes as a probe, the holonomy $V$ is an
arbitrary source, and we will put $V^{-1} \rightarrow V$.

Let us now follow this system through the geometric transition. The $N$
branes disappear, and the background geometry becomes the resolved
conifold. However, the $M$ probe branes are still there. The first
conjecture of Ooguri and Vafa is that these branes are wrapping a Lagrangian
submanifold ${\cal C}_{\cal K}$ of ${\cal O}(-1) \oplus {\cal O}(-1)
\rightarrow \IP^1$ that can be obtained from ${\widehat {\cal C}}_{\cal
K}$ through the geometric transition. The final outcome is therefore
the existence of
a map
\be
\label{ovmap}
\{ {\rm knots} \, \, {\rm in} \, \,  {\bf S}^3 \} \rightarrow \{  {\rm
Lagrangian} \,\,
{\rm submanifolds} \, \, {\rm in } \, {\cal O}(-1) \oplus {\cal O}(-1)
\rightarrow \IP^1 \}
\ee
which sends
\be
{\cal K} \rightarrow {\cal C}_{\cal K}.
\ee
Moreover, one has $b_1({\cal C}_{\cal K})=1$. This conjecture is clearly
well-motivated in the physics of the problem, and some
aspects of the map (\ref{ovmap}) are already well understood: in \cite{ov}
Ooguri and Vafa constructed ${\cal C}_{\cal K}$ explicitly when ${\cal K}$
is the unknot, and \cite{lmv} proposed Lagrangian submanifolds for certain
algebraic knots and links (including torus knots). Taubes has generalized
this proposal \cite{taubes} and constructed in detail a map from a wide class
of knots to Lagrangian submanifolds in the
resolved conifold. Later on we will discuss the
case of the unknot.

The resulting Lagrangian submanifold ${\cal C}_{\cal K}$ in the
resolved geometry provides boundary conditions for open strings,
and therefore it gives the open string sector that is needed in
order to extend the large $N$ duality to Wilson loops. The second
conjecture of \cite{ov} states that the free energy of open
topological strings (\ref{totalfreeopen}) with boundary conditions
specified by ${\cal C}_{\cal K}$ is identical to the free energy
of the deformed Chern-Simons theory with action (\ref{defact}),
which is nothing but (\ref{convevs}): \be F_{\rm string}(V)=F_{\rm
CS}(V). \label{wilsondual} \ee Notice that, since $b_1({\cal
C}_{\cal K})=1$, the topological sectors of maps with positive
winding numbers correspond to vectors $\vec k$ labelling the
connected vevs, and one finds
\begin{equation}
\label{freelog}
i^{|\vec k|} \sum_{g=0}^{\infty} F_{g, \vec k} (t) g_s ^{2g-2 + |\vec k|}
={1 \over \prod_j j^{k_j}} W_{\vec k}^{(c)}.
\label{concrewil}
\end{equation}
Of course, $F_{g, \vec k} (t)$ are (up to constants) the functions of the
't Hooft parameter that appeared in (\ref{opencon}). The variable
$\lambda$ defined in (\ref{varias}) that appears in the Chern-Simons
invariants of knots and links is related to the 't Hooft parameter through
$$
\lambda={\rm e}^t.
$$
Notice that the Chern-Simons invariants are labelled by vectors
$\vec k$, therefore they only give rise to positive winding
numbers in the string side. At the same time, they involve both
positive and negative powers of $\lambda$, while in the string
side we only have negative powers. Therefore, in order to make
(\ref{wilsondual}) precise, we further need some sort of analytic
continuation that gives an appropriate matching of the variables.
In the cases where both sides of the equality are known, there is
such an analytic continuation, and it is expected that this will
be the case in more general situations. Up to these subtleties,
(\ref{concrewil}) tells us that the Chern-Simons invariant in the
left-hand side is a generating function for open Gromov-Witten
invariants, for all degrees and genera, but with fixed boundary
data ({\it i.e.} the number of holes and the winding numbers). To
extract a particular open Gromov-Witten invariant from the
Chern-Simons invariant, we consider the connected vev labelled by
the vector $\vec k$ associated to the boundary data, we write it
in terms of $\lambda={\rm e}^t$ and $q={\rm e}^x$, and then we
expand the result in powers of $x=ig_s$. The coefficients of this
series, which are polynomials in $\lambda$, are then equated to
the generating function of open Gromov-Witten invariants at fixed
genus $g$.

We should mention that, although we have focused on knots
for simplicity,  all these
results can be extended to links, as shown in \cite{lmv}.

\subsection{BPS invariants for open strings from knot invariants}

In section 2 we have learned that
Gromov-Witten invariants can be written in terms of integer, or BPS
invariants. We will now find what is the precise relation between
Chern-Simons invariants and these integer invariants. This will lead
to some surprising structure results for the Chern-Simons invariants of knots.

The first step is to introduce the so-called
$f$-polynomials, through
the relation:
\begin{equation}
\label{wilop}
F_{\rm CS}(V)=\sum_{n=1}^\infty
\sum_R {1\over n} f_R(q^n, \lambda ^n) {\rm Tr}_R V^n.
\end{equation}
As shown in \cite{lm,lmqa}, the $f_R$ polynomials are completely 
determined by this equation, and can be expressed in terms of the
usual vevs of Wilson loops $W_R$ by:
\begin{eqnarray}
\label{explinver}
f_R (q, \lambda)&=& \sum_{d, m=1}^{\infty} (-1)^{m-1} {\mu (d) \over d m}
\sum_{\vec k_1, \cdots, \vec k_m} \sum_{R_1, \cdots, R_m} \chi_R (
C((\sum_{j=1}^l \vec k_j)_d)) \nonumber\\
&\times& \prod_{j=1}^m { \chi_{R_j}(C(\vec k_j)) \over z_{\vec k_j}}
W_{R_j}(q^d, \lambda^d),
\end{eqnarray}
where $\vec k_d$ is defined as follows: $(\vec k_d)_{di}=k_i$ and has zero
entries for the other components. Therefore, if $\vec k= (k_1, k_2, \cdots)$,
then
$$
\vec k_d =(0, \cdots, 0,k_1,0,\cdots, 0, k_2, 0,\cdots)$$ where
$k_1$ is in the $d$-th entry, $k_2$ is in the $2d$-th entry, and
so on. The sum over $\vec k_1, \cdots, \vec k_m$ is over all
vectors with $|\vec k_j| >0$. In (\ref{explinver}), $\mu (d)$
denotes the Moebius function. Recall that the Moebius function is
defined as follows: if $d$ has the prime decomposition
$d=\prod_{i=1}^a p_i^{m_i}$, then $\mu(d)=0$ if any of the $m_i$
is greater than one. If all $m_i=1$ ({\it i.e.} $d$ is
square-free) then $\mu(d)=(-1)^a$. Some examples of
(\ref{explinver}) are
\begin{eqnarray}
\label{examples}
f_{\tableau{1}}(q,\lambda)&=&W_{\tableau{1}}(q, \lambda),
\nonumber\\
f_{\tableau{2}}(q,\lambda)&=&W_{\tableau{2}}(q,\lambda)
-{1\over 2}\bigl( W_{\tableau{1}}(q,\lambda)^2+
W_{\tableau{1}}(q^2,\lambda^2)
\bigr),\nonumber\\
f_{\tableau{1 1}}(q, \lambda)&=&W_{\tableau{1 1}}(q,\lambda)
-{1\over 2}\bigl( W_{\tableau{1}}(q,\lambda)^2-
 W_{\tableau{1}}(q^2,\lambda^2)
\bigr).
\end{eqnarray}
Therefore, given a representation $R$ with $\ell$ boxes, the
polynomial $f_R$ is given by $W_R$, plus some ``lower order
corrections'' that involve $W_R'$ where $R'$ has $\ell'<\ell$
boxes. One can then easily compute these polynomials starting from
the results for vevs of Wilson loops in Chern-Simons theory.
Although we are calling $f_R$ polynomials, they are not, strictly
speaking. In fact, it follows from the multicovering/bubbling
formula that the $f_R$ have the structure \be \label{fpolstru} f_R
(q, \lambda)={P_R (q, \lambda) \over q^{1 \over 2} - q^{-{1 \over
2}}}. \ee But we can be more precise about the structure of $f_R$.
As shown in \cite{lmv}, one can write the $f_R$ in terms of even
more basic objects, that were denoted by $\widehat f_R$. The
precise relation between them is \be \label{fsrel} f_R =\sum_{R'}
M_{R R'} {\widehat f}_{R'} \ee where the sum in $R'$ runs over all
representations with the same number of boxes than $R$, and the
matrix $M_{R R'}$ is given by: \be \label{defm} M_{R
R'}=\sum_{R''} C_{R R' R''} S_{R''}(q). \ee In this equation,
$C_{R R' R''}$ are the Clebsch-Gordon coefficients of the
symmetric group. They can be explicitly written in terms of
characters \cite{fh}: \be C_{R\,R'\,R''} =\sum_{\vec k} {|C(\vec
k)| \over \ell !} \chi_R (C(\vec k))  \chi_{R'} (C(\vec k))
\chi_{R''} (C(\vec k)). \label{cgordon} \ee The $S_R(q)$ are
monomials defined as follows. If $R$ is a hook or L-shaped
representation of the form \be \tableau{6 1 1 1 1} \ee with $\ell$
boxes in total, and $\ell-d$ boxes in the first row, then \be S_R
(q)=(-1)^d q^{ -{\ell -1 \over 2}+d}, \ee and $S_R(q)=0$ for the
rest of the representations. For example, for the case of two
boxes one has that $S_{\tableau{2}}(q) =q^{-1/2}$ and
$S_{\tableau{1 1}}(q)=-q^{1/2}$, while for $\ell=3$ one has \be
S_{\tableau{3}}(q)=q^{-1}, \,\,\ S_{\tableau{2 1}}(q)=-1, \,\,\
S_{\tableau{1 1 1}}(q)=q.\ee The square matrix $M_{R R'}$ that
relates $f_R$ to $\widehat f_R$ is invertible. This can be easily
seen: define the polynomials $P_{\vec k}(q)$, labelled by
conjugacy classes, as the character transforms of the monomials
$S_R (q)$: \be\label{pk} P_{\vec k}(q)=\sum_R \chi_R(C(\vec k))
S_R(q).\ee It can be seen that \be \label{Pex} P_{\vec k}(q)={
\prod_j (q^{-{j\over 2}} -q^{j\over 2})^{k_j} \over q^{-{1\over
2}} -q^{1\over 2}}. \ee In terms of these polynomials, the matrix
$M_{R R'}$ is written as \be M_{R R'}= \sum_{{\vec k}}
 {1 \over z_{\vec k}}
\chi_{R}(C({\vec k})) \chi_{R'}
(C({\vec k}))P_{{\vec k}}(q),
\ee
and using the orthogonality of the characters one can see that
\be
M_{R R'}^{-1}=\sum_{{\vec k}}
 {1 \over z_{\vec k}}
\chi_{R}(C({\vec k})) \chi_{R'}
(C({\vec k})) (1/P_{{\vec k}}(q)).
\ee
Therefore, one can obtain the polynomials $\widehat f_R$ from the
$f_R$, {\it i.e.} one can obtain the polynomials $\widehat
f_R$ from the knot
invariants of Chern-Simons theory. The claim is now that the $\widehat f_R$ are
generating functions for the BPS invariants $N_{R,g,Q}$ that were
introduced in (\ref{openBPS}). More precisely, one has
\be
\label{BPSexp}
 {\widehat f}_{R}(q, \lambda)=
\sum_{g \ge 0}\sum_Q N_{R,g,Q}
(q^{-{1\over 2}}-q^{1 \over 2})^{2g-1}\lambda^Q
\ee
Therefore, this gives a very precise way to compute the BPS invariants
$N_{R,g,Q}$ from Chern-Simons theory: compute the usual vevs $W_R$,
extract $f_R$ through the relation (\ref{explinver}), compute $\widehat
f_R$, and expand them as in (\ref{BPSexp}).

We would like to point out two important things. First, the fact
that one can extract the integer invariants $N_{R,g,Q}$ from Chern-Simons
theory in the way we have just described is by no means obvious and
constitutes a strong check of the large $N$
duality between Chern-Simons theory and topological strings.
We will see examples of this in the next subsection. Another important
comment is that the statement that $\widehat f_R$ have the structure
predicted in (\ref{BPSexp}) is equivalent to the multicovering/bubbling
formula for open string invariants (\ref{multopen}) (more precisely, it is
equivalent to the strong version of this formula, which says that in
addition to
(\ref{multopen}) one can write the $n_{\vec k, g, Q}$ in terms of integer
$N_{R,g, Q}$ through (\ref{openBPS})). This is easily seen by noticing
that, according to (\ref{fsrel}) and (\ref{defm}), $f_R$ is given by
\be
\label{frcom}
f_{R}(q, \lambda)=
\sum_{g\ge 0} \sum_{Q}
\sum_{R', R''}
C_{R\,R'\,R''}S_{R'}(q)
N_{R'',g,Q}
 (q^{-{1\over 2}}-q^{1\over 2})^{2g-1}\lambda^Q.
\ee
If we now write the exponent in the r.h.s. of
(\ref{wilop}) in the $\vec k$ basis, it is easy to see that one obtains
precisely (\ref{multopen}), after making use of (\ref{Pex}).

The physical origin of the structure of $f_R$ (and therefore of
the multicovering/bubbling formula for open Gromov-Witten
invariants) can be easily understood in physical terms. We will
give a short account, referring the reader to \cite{lmv} for more
details. In the D-brane approach to open string instantons, one
regards the open Riemann surfaces ending on a Lagrangian
submanifold as D2-branes ending on $M$ D4-branes wrapping the
Lagrangian submanifold. Following the approach of \cite{gv}, we
have to study the moduli space of D2-branes ending on D4-branes.
This moduli space is the product of three factors: the moduli of
Abelian gauge fields on the worldvolume of the D2 brane, the
moduli of geometric deformations of the D2's in the ambient space,
and finally the Chan-Paton factors associated to the boundaries of
the D2 which appear in the D4 as magnetic charges \cite{ov}. If
the D2's are genus $g$ surfaces with $\ell$ holes in the relative
cohomology class labelled by $Q$, the moduli space of Abelian
gauge fields gives rise to the Jacobian $J_{g,\ell}={\bf
T}^{2g+\ell-1}$, and the moduli of geometric deformations will be
a manifold ${\cal M}_{g,\ell, Q}$. Finally, for the Chan-Paton
degrees of freedom we get a factor of $F$ (the fundamental
representation of $SU(M)$) from each hole. The Hilbert space is
obtained by computing the cohomology of these moduli, and we
obtain \be F^{\otimes \ell} \otimes H^*(J_{g,\ell}) \otimes
H^*({\cal M}_{g,\ell, Q}). \ee An important point is that this
Hilbert space is associated with the moduli space of $\ell$ {\it
distinguished} holes, which is not physical, and we have to mod
out by the action of the permutation group $S_{\ell}$. We can
factor out the cohomology of the Jacobian ${\bf T}^{2g}$ of the
``bulk'' Riemann surface, $H^*({\bf T}^{2g})$, since the
permutation group does not act on it. The projection onto the
symmetric piece can be easily done using the Clebsch-Gordon
coefficients $C_{R\,R'\,R"}$ of the permutation group $S_{\ell}$
\cite{fh}: \ben \label{hilbertdec} & & {\rm Sym}\Bigl(F^{\otimes
\ell} \otimes H^*(({\bf S}^1)^{\ell-1})
\otimes H^*({\cal M}_{g,\ell, Q})\Bigr)=\nonumber\\
\,\,\,\,\,\, & &  \sum_{R\,R'\,R''} C_{R\,R' \, R''} {\bf
S}_R(F^{\otimes \ell})\otimes{\bf S}_{R'}(H^*(({\bf
S}^1)^{\ell-1})) \otimes {\bf S}_{R''}(H^*({\cal M}_{g,\ell,Q}))
\een where ${\bf S}_R$ is the Schur functor that projects onto
the corresponding subspace. The space ${\bf S}_R(F^{\otimes
\ell})$ is nothing but the vector space underlying the irreducible
representation $R$ of $SU(M)$. ${\bf S}_{R'}(H^*(({\bf
S}^1)^{\ell-1}))$ gives the hook Young tableau, and the Euler
characteristic of ${\bf S}_{R''}(H^*({\cal M}_{g,\ell, Q}))$ is
the integer invariant $N_{R'',g,Q}$. Therefore, the above
decomposition corresponds very precisely to (\ref{frcom}).

All the results above have been stated for knot invariants in the
canonical framing. The situation for arbitrary framing was
analyzed in detail in \cite{mv}. Suppose that we consider a knot
in ${\bf S}^3$ in the framing labelled by an integer $p$ (the
canonical framing corresponds to $p=0$). Then, the integer
invariants $N_{R,g,Q}(p)$ are obtained from (\ref{explinver}) but
with the vevs
\begin{equation}
\label{framefin}
W^{(p)}_R (q, \lambda) = (-1)^{\ell p} q^{ {1\over 2} p \kappa_R}
W_R (q, \lambda),
\end{equation}
where $\kappa_R$ is defined in (\ref{kapar}). One has, for example,
\begin{eqnarray}
\label{pexamples}
f^{(p)}_{\tableau{1}}(q,\lambda)&=&(-1)^pW_{\tableau{1}}(q, \lambda),
\nonumber\\
f^{(p)}_{\tableau{2}}(q,\lambda)&=&q^p W_{\tableau{2}}(q,\lambda)
-{1\over 2}\bigl( W_{\tableau{1}}(q,\lambda)^2+ (-1)^p
W_{\tableau{1}}(q^2,\lambda^2)
\bigr),\nonumber\\
f^{(p)}_{\tableau{1 1}}(q, \lambda)&=&q^{-p}W_{\tableau{1 1}}(q,\lambda)
-{1\over 2}\bigl( W_{\tableau{1}}(q,\lambda)^2-(-1)^p
 W_{\tableau{1}}(q^2,\lambda^2)
\bigr),
\end{eqnarray}
and so on. Notice that the right framing factor in order to match
the topological string theory prediction is (\ref{explcas}), and not
(\ref{explcasun}).
This is yet another indication that the duality of \cite{gv} involves the
$U(N)$ gauge group, not the $SU(N)$ group. The rationale for introducing
the extra sign $(-1)^p$ is not completely clear in the context of
Chern-Simons theory, and it was introduced by consistency with the results
for the B-model in \cite{akv}. This sign is crucial for integrality of
$N_{R,g,Q}(p)$.

All the above results on $f$-polynomials, integer invariant structure,
etc., can be extended to links, see \cite{lmv,lmqa}.

\subsection{Tests involving Wilson loops}
There are two types of tests of the large $N$ duality involving Wilson
loops: a test in the strong sense, in which one verifies that the open
Gromov-Witten invariants agree with the Chern-Simons amplitude, and a test
in the weak sense, in which one verifies that the Chern-Simons knot
invariants satisfy the integrality properties that follow from the
conjectured dual description.

\begin{figure}
\begin{center}
\epsfysize=7cm
\epsfbox{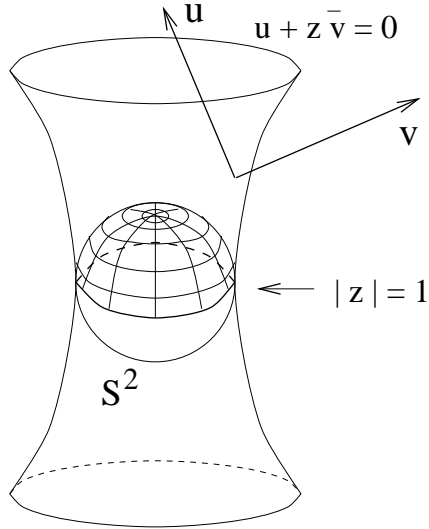}
\end{center}
\label{unknot}
\caption{This figure \cite{ov} represents the Lagrangian submanifold in
 ${\cal O}(-1) \oplus
{\cal O}(-1) \rightarrow \IP^1$ that corresponds to the unknot in ${\bf
S}^3$. The notation is as in \cite{ov}, and is related to ours by
$u\rightarrow x$ and $v\rightarrow -u$.}
\end{figure}

The only test so far of the duality in the strong sense is for the framed
unknot. In this case, we know both sides of the duality in detail and we
can compare the results. Let us start with the string description. The first
thing we need is a construction of the Lagrangian submanifold ${\cal
C}_{\cal K}$ that corresponds to the unknot in ${\bf S}^3$. This was done
by Ooguri and Vafa in \cite{ov}. The construction goes as follows.
Let us start with $T^* {\bf S}^3$
expressed as (\ref{defconifold}) ,
and consider the following anti-holomorphic involution on it.
\be\label{ztwo} \eta_{1,2} = \bar{\eta}_{1,2},
~~\eta_{3,4} = - \bar{\eta}_{3,4}.
\ee
The symplectic form $\omega$ changes its sign under
the involution, therefore its fixed point set is a
Lagrangian submanifold of $T^* {\bf S}^3$.
If we write $\eta_{\mu} = x_\mu + i p_\mu$, the invariant locus
of the action (\ref{ztwo}) is
\be \label{fixed} p_{1,2} = 0,~~ x_{3,4} = 0
\ee
and intersects the deformed conifold at
\be
\label{fixedmore}
x_1^2 + x_2^2 = a + p_3^2 + p_4^2.
\ee
Therefore, the fixed point locus intersects ${\bf S}^3$ along
the equator, which is an unknot described by the equations
$$x_1^2 + x_2^2 = a, \quad x_3=x_4=0.$$
We conclude that, if we denote by ${\cal U}$ the unknot in ${\bf S}^3$, the
above fixed point locus defined by (\ref{ztwo}) is the Lagrangian
submanifold ${\widehat C}_{\cal U}$. Now we want to construct the
Lagrangian submanifold ${\cal C}_{\cal U}$, obtained from
${\widehat C}_{\cal U}$ after the conifold
transition. To do that, we continue to identify it with the invariant
locus of the anti-holomorphic
involution. We can describe this explicitly by using the coordinates
 $(x,u,z)$ or
$(y,v, z^{-1})$ defined in (\ref{newcoords})
and (\ref{rescon}). In these coordinates,
${\widehat C}_{\cal U}$ is characterized by
\be
\label{aftertransition}
x = \bar{y}, ~~u = \bar{v},\ee
and the conifold equation (\ref{conifold}) restricted to
${\widehat {\cal
C}}_{\cal U}$ becomes
\be
\label{samesize} x \bar x = u \bar u. \ee
The complex coordinate on the base $\IP^1$ defined by (\ref{rescon})
is
\be \label{phase} z ={x \over \bar{u}},
\ee
but since $|x|=|u|$, $z$ is a phase. We then find that ${\cal C}_{\cal U}$
is a line bundle over the equator $|z| = 1$ of $\IP^1$, and
the fiber over $z$ is the subspace of ${\cal O}(-1)+{\cal O}(-1)$ given by
$x =z\bar{u}$ (remember that $x$, $u$ are complex coordinates for the fibers).
In particular, ${\cal C}_{\cal U}$ intersects with the $\IP^1$ at the base
along $|z|=1$, see \figref{unknot}.

The open Gromov-Witten invariants associated to open strings in
${\cal O}(-1) \oplus {\cal O}(-1) \rightarrow \IP^1$ whose
boundaries end in the above Lagrangian submanifold have been
computed in \cite{kl,song,mayr} (see also \cite{gz}). The
procedure relies on localization formulae, as in the closed string
case. However, in the open string case, it has been realized that
the open invariants depend on an extra choice of an integer (the
calculation depends on the weights on the localizing torus
action). This is precisely the dependence we expect on
Chern-Simons theory, since there is a choice of framing also
labelled by an integer. This framing ambiguity in the context of
open strings was first discovered in the B-model \cite{akv}, and
subsequently confirmed in the A-model computation of \cite{kl} as
well as in other examples \cite{gz,mayr}. Let us now make a
detailed comparison of the answers. Katz and Liu \cite{kl} compute
the open Gromov-Witten invariants $F_{w, g}^Q$ for $Q=\ell/2$,
where $\ell=\sum_i w_i$, and obtain:
\begin{eqnarray}
\label{klform}
F_{w, g}^{\ell/2}&=& (-1)^{p\ell+1}
(p(p+1))^{h-1}\biggl( \prod_{i=1}^h { \prod_{j=1}^{w_i-1}
(j+w_i p) \over (w_i-1)! } \biggr)
\nonumber\\
& & \cdot {\rm Res}_{u=0}
\int_{{\overline M}_{g,h}}
{c_g (\DE ^{\vee}(u))c_g(\DE ^{\vee} ((-p-1)u)) c_g (\DE ^{\vee} (p u))
u^{2h-4} \over \prod_{i=1}^h (u- w_i \psi_i)}.
\end{eqnarray}
In this formula, $\overline M_{g,h}$ is the Deligne-Mumford moduli space
of genus $g$ stable curves with $h$ marked points,
$\DE$ is the Hodge bundle over $\overline M_{g,h}$, and its dual
is denoted by $\DE ^{\vee} $. The Chern classes of the
Hodge bundle will be denoted by:
\begin{equation}\label{chodge}
\lambda_j =c_j (\DE).
\end{equation}
In (\ref{klform}), we have written
\begin{equation}\label{serieshod}
c_g(\DE ^{\vee} (u))= \sum_{i=0}^g c_{g-i} (\DE ^{\vee}) u^i,\end{equation}
and similarly for the other two factors. The integral in
(\ref{klform}) also
involves the $\psi_i$ classes of two-dimensional topological gravity,
which are constructed as follows. We first define the
line bundle ${\mathcal L}_i$ over ${\overline M}_{g,h}$ to be the line
bundle whose fiber over each stable curve $\Sigma$ is the cotangent space
of $\Sigma$ at $x_i$ (where $x_i$ is the $i$-th marked point). We then have,
\begin{equation}
\label{psiclas}
\psi_i =c_1 ({\mathcal L}_i),\,\,\,\ i=1, \cdots, h.\end{equation}
The integrals of the $\psi$ classes can be obtained by the
results of Witten and Kontsevich on 2d topological gravity
\cite{wtwodg,k}, while the integrals involving
$\psi$ and $\lambda$ classes (the so-called Hodge integrals) can be in
principle computed by reducing them to pure $\psi$ integrals \cite{faber}.
Explicit formulae for some Hodge integrals can be found in \cite{gp}.

In the above formula (\ref{klform}), $p$ is an integer that
parameterizes the ambiguity in the open string calculation. A
particularly simple case of the above expression is when $p=0$,
{\it i.e.} the standard framing. The only contribution comes from
$h=1$, and the above integral boils down to \be {\rm Res}_{u=0}
\int_{{\overline M}_{g,1}} {\lambda_g c_g (\DE ^{\vee}(u))c_g(\DE
^{\vee} (-u)) u^{2h-4} \over (u- w \psi_1)}, \ee where $w$ is the
winding number. The Mumford relations \cite{mumford} give
$c(\DE)c(\DE^{\vee})=1$, which implies \be c_g(\DE
^{\vee}(u))c_g(\DE ^{\vee} (-u))=(-1)^g u^{2g} \ee After taking
the residue, we end up with the following expression for the open
Gromov-Witten invariant: \be F_{w,
g}^{w/2}=-w^{2g-2}\int_{{\overline M}_{g,1}} \psi_1^{2g-2}
\lambda_g. \ee The above Hodge integral has been computed in
\cite{fp}, and it is given by $b_g$, where $b_g$ is defined by the
generating functional \be \sum_{g=0}^{\infty} b_g x^g = { x/2
\over \sin (x/2)}. \ee We can now sum over all genera and all
positive winding numbers to obtain \cite{kl} \be
F(V)=-\sum_{d=1}^{\infty}{ { \rm e}^{dt/2} \over 2 d \sin \Bigl(
{d g_s \over 2} \Bigr)}{\rm Tr} V^d. \label{klanswer} \ee Notice
that the above open Gromov-Witten invariants correspond to a disk
instanton wrapping the northern hemisphere of $\IP^1$, with its
boundary on the equator, together with all the multicoverings and
bubblings at genus $g$ \footnote{In this equation we have chosen
the sign for the instantons wrapping the northern hemisphere in
such a way that one has ${\rm e}^{dt}$ in the generating function,
in \ order to compare to the results in \cite{mv}.}. Let us now
compare to the Chern-Simons computation. In the case of the unknot
in the canonical framing, Ooguri and Vafa showed \cite{ov} that
the generating function (\ref{convevs}) can be explicitly computed
to all orders. The reason is that the quantum dimension in the
representation $R$ can be regarded as the trace in the
representation $R$ of an $N \times N$ diagonal matrix $U_0$ whose
$i$-th diagonal entry is \be \exp \Bigl[ - {\pi i \over k+N}
(N-2i-1) \Bigr]. \ee This is easily seen by remembering that
$\rho$ lives in the dual of the Cartan subalgebra $H$, and by
using the natural isomorphism between $H$ and $H^*$ induced by the
Killing form we obtain the above result from (\ref{qdimchar}).
Notice that $U_0$ is like a ``master field'' that gives the right
answer by evaluating a ``classical'' trace. Therefore, one can
compute $F_{\rm CS}(V)$ by substituting ${\rm Tr}U_0^n$ in
(\ref{ovop}), to obtain \be F(V)=-i \sum_{d=1}^{\infty} { {\rm
e}^{dt/2} - {\rm e}^{-dt/2} \over 2d \sin \Bigl( {d g_s \over 2}
\Bigr)} {\rm Tr}V^{d}. \ee The answer from Chern-Simons theory
contains the contribution given in (\ref{klanswer}), together with
a similar contribution (with ${\rm e}^{t/2}$) that corresponds to
holomorphic maps wrapping the southern hemisphere of the ${\bf
P}^1$.

What happens for $p \not=0$? In that case, it is no longer possible to
sum up all the correlation functions, but we can still compute the
connected vevs $W_{\vec k}^{(c)}$ at arbitrary framing \cite{mv}.
To do that, remember that the $W_R$ for the unknot in the canonical framing are
just the quantum dimensions of the representation $R$ given in
(\ref{expf}). We have to correct them with the framing factor as prescribed
in (\ref{framefin}), compute the $W_{\vec k}$ with
Frobenius formula, and then extract the connected piece by using:
\begin{equation}
\label{connew}
{1 \over z_{\vec k}}W_{\vec k}^{(c)} =\sum_{n\ge 1} {(-1)^{n-1} \over n}
\sum_{\vec k_1, \cdots, \vec k_n} \delta_{\sum_{i=1}^n \vec k_i, \vec k}
\prod_{i=1}^n {W_{\vec k_i} \over z_{\vec k_i}}.
\end{equation}
In this equation, the second sum is over $n$ vectors
$\vec k_1, \cdots, \vec k_n$
such that $\sum_{i=1}^n \vec k_i =\vec k$ (as indicated by
the Kronecker delta),
and therefore the right hand side of (\ref{connew}) involves a
finite number of terms. The generating functional for the open Gromov-Witten
invariants is then explicitly given by
\begin{eqnarray}
\label{moreor}
& &\sum_{Q}\sum_{g=0}^\infty F_{\vec k, g}^{Q} g_s^{2g-2 + |\vec k|}
{\rm e}^{Qt}
=\nonumber\\
& & (-1)^{p \ell} i^{-|\vec k| -\ell} \prod_j k_j! \sum_{n\ge 1}
{(-1)^{n} \over n}
\sum_{\vec k_1, \cdots, \vec k_n} \delta_{\sum_{\sigma=1}^n \vec k_\sigma, \vec k}
\sum_{R_{\sigma}} \prod_{\sigma =1}^n { \chi_{R_{\sigma}} (C(\vec k_{\sigma}))
\over z_{\vec k_{\sigma}} }\nonumber\\ & &\cdot
{\rm e}^{i p \kappa_{R_{\sigma}}g_s/2}
\prod_{ 1\le i < j \le c_{R_{\sigma}}}
{\sin \Bigl[ (l^{\sigma}_i -l^{\sigma}_j +j-i) g_s/2\Bigr]
\over \sin \Bigl[ (j-i) g_s /2\Bigr]}
\prod_{i=1}^{c_{R_{\sigma}}} {\prod_{v=-i+1}^{l_i^{\sigma}-i}
\bigl(
{\rm e}^{ {t\over 2} + {iv g_s\over 2}}
-{\rm e}^{-{t\over 2}-{iv g_s\over 2} }\bigr)
\over
\prod_{v=1}^{l^{\sigma}_i} 2 \sin
\Bigl[ (v-i+c_{R_{\sigma}}) g_s /2\Bigr]}.\nonumber\\
\end{eqnarray}
Let us compare this expression with the result of Katz and Liu in some
simple examples with $h=1$. Notice that the Chern-Simons result is slightly
more general, since it gives the answer for any $Q$, while (\ref{klform})
only computes $Q=\ell/2$. For Riemann surfaces with one hole
the homotopy class of the map is given by a single winding number $w$.
For $g=1$, one finds from (\ref{klform}):
 \begin{equation}
\label{resultone}
F_{w,1}^{w/2} ={(-1)^{pw}\over (w-1)!} \prod_{l=1}^{w-1} (l + wp) \Biggl(
\biggl(  \int_{ {\overline M}_{1,1}} \lambda_1-w \psi_1
\biggr) p(p+1) + \int_{ {\overline M}_{1,1}} \lambda_1 \Biggr),
\end{equation}
and for $g=2$,
\begin{eqnarray}
\label{resultwo}
F_{w,2}^{w/2} &=&{(-1)^{pw}\over (w-1)!} \prod_{l=1}^{w-1} (l + wp) \Biggl(
\biggl(\int_{ {\overline M}_{2,1}} w^2 \psi_1^4 -
w \psi_1^3 \lambda_1 + \psi_1^2 \lambda_2  \biggr) w^2 p^3 (p+2) \nonumber\\
&+& \biggl(\int_{ {\overline M}_{2,1}} w^3 \psi_1^4 -
2 w^2 \psi_1^3 \lambda_1 -
\psi_1\lambda_1\lambda_2 + 3w \psi_1^2\lambda_2  \biggr)
wp^2 \nonumber\\
& +&  \biggl(\int_{ {\overline M}_{2,1}} -w^2 \psi_1^3 \lambda_1 -
\psi_1\lambda_1 \lambda_2 + 2w \psi_1^2\lambda_2  \biggr)
wp  + w^2 \int_{ {\overline M}_{2,1}} \psi_1^2\lambda_2\Biggr).
\end{eqnarray}
To obtain this expression, we have used the Mumford relation,
which implies in particular $\lambda_2^2=0$
and $\lambda_1^2 =2\lambda_2$. On the other hand, the Chern-Simons answer
for the connected vevs when $w=1$ and $w=2$ is:
\begin{eqnarray}\label{computa}
iW_1^{(c)} (g_s) &= &{(-1)^p \over g_s}\biggl( 1 + {1 \over 24} g_s^2 +
{7 \over 5760} g_s^4+ {\mathcal O}(g_s^6)\biggr),\nonumber\\
{i\over 2} W_2^{(c)}(g_s)& = & { 1+ 2p
\over g_s}  \biggl( {1 \over 4}
- {1 \over 24} (p^2 +p - 1) g_s^2 \nonumber\\ & & \,\,\  +
{1 \over 1440}(7-11p -8p^2 + 6p^3 + 3p^4)g_s^4
+ {\mathcal O}(g_s^6)\biggr),
\end{eqnarray}
and so on. By using now the following values of the Hodge integrals for
$g=1$ \begin{equation}
\label{genone}
\int_{ {\overline M}_{1,1}}\psi_1=
\int_{ {\overline M}_{1,1}}\lambda_1={1\over 24}
\end{equation}
and for $g=2$
$$
\int_{ {\overline M}_{2,1}}\psi_1^4={1\over 1152},\,\,\,\
\int_{ {\overline M}_{2,1}}\psi_1^3\lambda_1 ={1 \over 480}, $$
$$
\int_{ {\overline M}_{2,1}}\psi_1^2\lambda_2 ={7\over 5760},\,\,\,\
\int_{ {\overline M}_{2,1}}\psi_1\lambda_1\lambda_2 ={1 \over 2880},$$
we find perfect agreement between (\ref{resultone}) and (\ref{resultwo})
for $w=1,2$, and the Chern-Simons answer. Moreover, it is
in principle possible to compute all the integrals over
${\overline M}_{g,h}$ that appear in (\ref{klform}) from the explicit
expression (\ref{moreor}). These Hodge integrals include an arbitrary number
of $\psi$ classes and up to three $\lambda$ classes. Therefore, all
correlation functions of two-dimensional topological gravity can in principle
be extracted from (\ref{moreor}). It should be noted, however, that some
of the simple structural properties of (\ref{klform})
are not at all obvious from
(\ref{moreor}). For example, for $g=0$, $h=1$, (\ref{klform}) gives a fairly
compact expression for the open Gromov-Witten invariant, and the fact
that this equals the Chern-Simons answer amounts to a rather
nontrivial combinatorial identity. It is also possible to check that the
open Gromov-Witten invariants obtained in this way can be expressed in
terms of BPS invariants, see \cite{mv} for more details.

Unfortunately, although there are proposals for the Lagrangian submanifolds
that should correspond to other knots \cite{lmv, taubes}, the associated
open Gromov-Witten invariants have not been computed yet, so one is forced
to test the conjecture in the ``weak'' sense of showing that one can
extract integer
invariants from the Chern-Simons invariants in the way described
before. This was done in \cite{lm,rs,lmv} for various knots and links and
it was shown in all cases that indeed such invariants can be extracted in a
highly nontrivial way. We will give a simple example of this, involving the
trefoil knot. By using the known values for the Chern-Simons invariants
(\ref{trefoil}), and the defining relations for the $f$-polynomials
(\ref{examples}), one can easily obtain:
\ben
\label{ftrefoil}
f_{\tableau{2}}(q,\lambda)
&=&{ q^{-{1\over 2}}{\lambda}( {\lambda}-1) ^2 \,\,( 1 + {q^2}) \,
     ( q + {{{\lambda}}^2}\,q - {\lambda}\,( 1 + {q^2} ))
\over q^{{1\over 2}} - q^{-{1\over 2}} },
\nonumber\\
f_{\tableau{1 1}}(q,\lambda)& =& - {1 \over q^3}f_{\tableau{2}}(q,\lambda).
\een
Notice that, although the Chern-Simons invariants have complicated
denominators, the $f$-polynomials have indeed the structure
(\ref{fpolstru}). One can go further and extract the BPS invariants
$N_{\tableau{2},g,Q}$, $N_{\tableau{1 1},g,Q}$ from (\ref{ftrefoil}), by
using (\ref{fsrel}). The results are presented in Table \ref{SymTa} and
Table \ref{AntTa}, respectively.
\begin{table}
\begin{center}
\begin{tabular}{|cccccc|}
 \hline
 $g$&$Q=1$ & 2&3&4&5  \\  \hline
 0&-2&8&-12&8&-2\\ \hline
 1&-1&6&-10&6&-1\\ \hline
2&0&1&-2&1&0\\ \hline
\end{tabular}
\end{center}
\caption{BPS invariants
for the trefoil knot in the symmetric representation.}
\label{SymTa}
\end{table}
\begin{table}
\begin{center}
\begin{tabular}{|cccccc|}
 \hline
 $g$&$Q=1$ & 2&3&4&5  \\  \hline
 0&-4
&16
&-24
&16
&-4\\ \hline
1&-4
&20
&-32
&20
&-4\\ \hline
2&-1
&8
&-14
&8
&-1\\ \hline
3
&0
&1
&-2
&1
&0 \\ \hline
\end{tabular}
\end{center}
\caption{BPS invariants
for the trefoil knot in the antisymmetric representation.}
\label{AntTa}
\end{table}
The above results have been obtained in the canonical framing. Some integer
invariants for the trefoil knot in arbitrary framing are listed in
\cite{mv}. Results for the BPS invariants of other knots and links
can be found in \cite{lmv}.

\sectiono{Large $N$ transitions and toric geometry}

The duality between Chern-Simons on ${\bf S}^3$ and closed topological
strings on the resolved conifold gives a surprising point of view on
Chern-Simons invariants of knots and links. However, from the ``gravity''
point of view we do not learn much about the closed string geometry, since
the resolved conifold is quite simple (remember that it only has one
nontrivial Gopakumar-Vafa invariant). It would be very interesting to find
a topological gauge theory dual to more complicated geometries, in such a
way that we could use our knowledge of the gauge theory side to learn about
enumerative invariants of closed strings, and about closed strings in
general.

Such a program was started by Aganagic and Vafa in \cite{av}. Their basic
idea is to construct geometries that locally contain $T^*{\bf S}^3$'s, and then follow
geometric transitions to dual geometries where the ``local''
deformed conifolds are
replaced by resolved conifolds. Remarkably, a large class of non-compact
toric manifolds can be realized in this way, as it was made clear in
\cite{amv}. Let us consider in detail an example that allows one to recover
the local $\IP^2$ geometry.

\begin{figure}
\leavevmode
\begin{center}
\epsfysize=6cm
\epsfbox{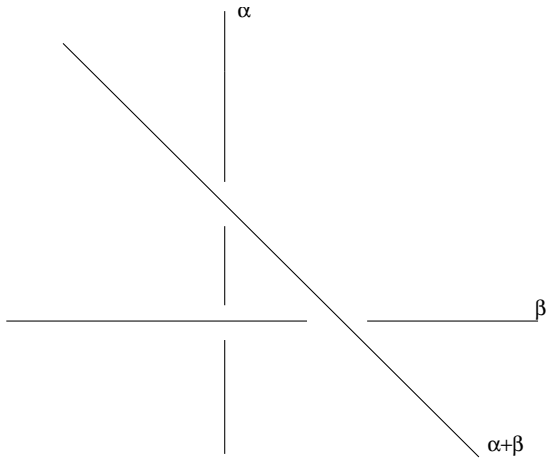}
\end{center}
\caption{This shows a Calabi-Yau which is a
${\bf T}^2 \times \IR$ fibration of $\IR^3$, where the
$\alpha$, $\beta$, and $\alpha+\beta$ cycles of the torus
degenerate at three lines.}
\label{p2defor} \end{figure}

\begin{figure}
\leavevmode
\begin{center}
\epsfysize=5cm
\epsfbox{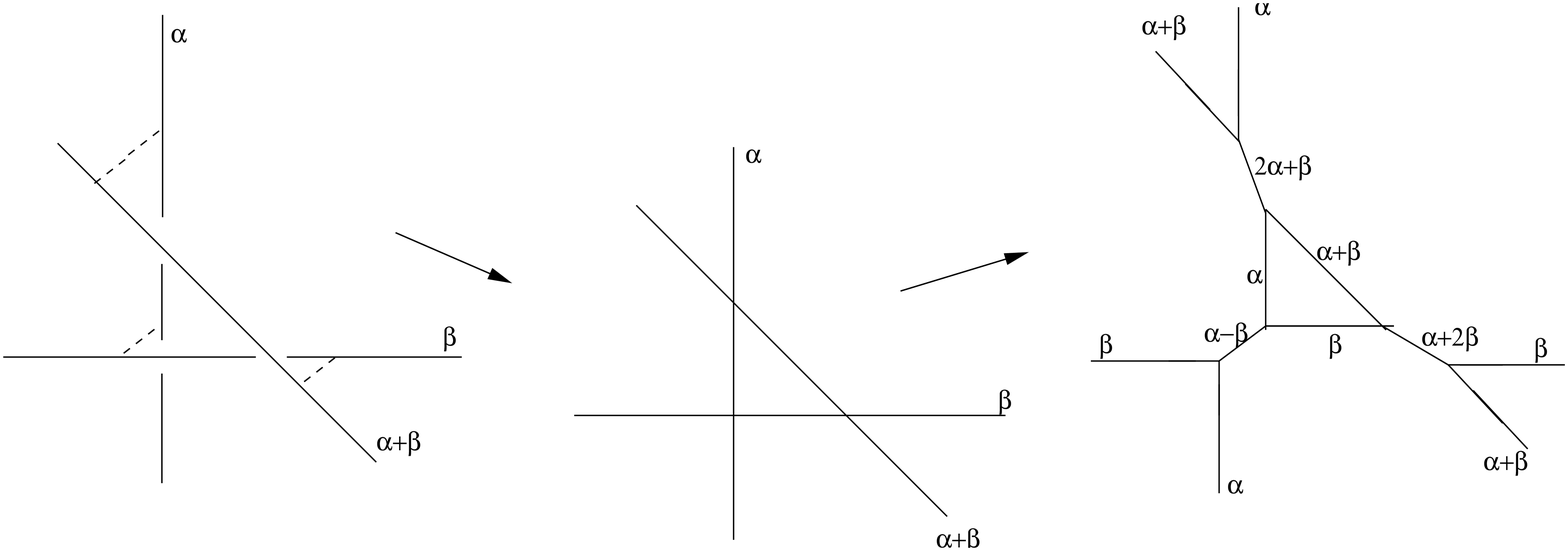}
\end{center}
\caption{This shows the geometric transition of the
Calabi-Yau in the previous figure.
In the leftmost geometry
there are three minimal 3-cycles.
The lengths of the
dashed lines are proportional to their sizes.
The intermediate geometry
is singular, and the figure on the right is the base of the smooth
toric Calabi-Yau after the transition. This Calabi-Yau
is related to $\IB_3$ by
flopping three $\IP^1$'s.}\label{threetrans}
\end{figure}

Recall from our discussion in section 4 that the deformed conifold can be
represented by a graph where one indicates the degeneration loci of the
cycles of the torus fiber. Following this idea, one can construct more
general ${\bf
T}^2 \times \IR$ fibrations of ${\bf R}^3$ by specifying degeneration loci
in the basis. An example of this is shown in \figref{p2defor}.
Notice that this geometry contains three ${\bf S}^3$'s, represented as
dashed lines in \figref{threetrans}. One can then think about a geometric
transition where the three-spheres go to zero size, and then the
corresponding singularities are blown-up to give a resolved geometry, as
shown in \figref{threetrans}. The resolved geometry turns out to be toric, and
in fact it can be obtained by three blowups of the Calabi-Yau manifold
${\cal O}(-3) \rightarrow \IP^2$. Up to flops of the three $\IP^1$'s,
the resulting geometry is the noncompact Calabi-Yau manifold given by the
del Pezzo surface $\IB_3$ together with its canonical bundle. To recover
the local $\IP^2$ geometry, one just sends the sizes of the three $\IP^1$'s
to infinity. The remaining ``triangle'' is the toric diagram for
the local $\IP^2$ geometry, see \cite{lv,akv}.

Let us now wrap $N_i$ branes, $i=1,2,3$, around the three ${\bf
S}^3$'s of the deformed geometry depicted in \figref{p2defor}.
What is the effective topological action describing the resulting
open strings? For open strings with both ends on the same ${\bf
S}^3$, the dynamics is described by Chern-Simons theory with gauge
group $U(N_i)$, therefore we will have three Chern-Simons theories
with groups $U(N_1)$, $U(N_2)$ and $U(N_3)$. However, there are
new sectors of open strings stretched between two spheres, giving
the nondegenerate instantons that we described in 4.2, following
\cite{csts}. Instead of describing these open strings in geometric
terms, it is better to use the spacetime physics associated to
these strings. In fact, a similar situation was considered when we
analyzed the incorporation of Wilson loops in the large $N$
duality. There we had two sets of intersecting D-branes, giving a
massless complex scalar field living in the intersection and in
the bifundamental representation of the gauge groups. Now, if we
focus, say, on the $N_1$, $N_2$ branes, we will get again a
complex scalar $\phi$ in $(N_1,\overline N_2)$. This complex
scalar is generically massive, and its mass is proportional to the
``distance'' between the two three-spheres, and it is given by a
complexified K\"ahler parameter that will be denote by $r$. We can
now integrate out this complex scalar field to obtain the
correction to the Chern-Simons actions on the three-spheres due to
the presence of the new sector of open strings, which is given by:
\ben \label{massiveov}
 {\cal O}(U_1, U_2; r) &=&\exp\Bigl[-{\rm Tr} \; \log(
{\rm e}^{r/2} U_1^{-1/2}\otimes
U_2^{1/2} - {\rm e}^{-r/2} U_1^{1/2}\otimes  U_2^{-1/2})\Bigr]\nonumber\\
&=&\exp\Bigl\{\sum_{n=1}^{\infty}\; \frac{{\rm e}^{-nr}}{n}
\;{\rm Tr} U_1^{n}\; {\rm Tr}U_2^{-n}\Bigr\},
\een
where $U_{1,2}$ are the holonomies of the corresponding
gauge fields around a loop. Note that the operator ${\cal O}$ is
the amplitude for a primitive annulus of size $r$ together with its
multicovers, as one can see from the first equation of (\ref{multiopen})
for $h=2$. This annulus ``connects''
the two ${\bf S}^3$'s, {\it i.e.} one of its boundaries is in one
three-sphere, and the other boundary is in the other sphere. The exponent
in (\ref{massiveov}) is the contribution
to $F_{\rm ndg}$ in (\ref{nondeg}) due to
these configurations of open strings, and $r$ is
the complexified area of the annulus.

\begin{figure}
\leavevmode
\begin{center}
\epsfysize=8cm
\epsfbox{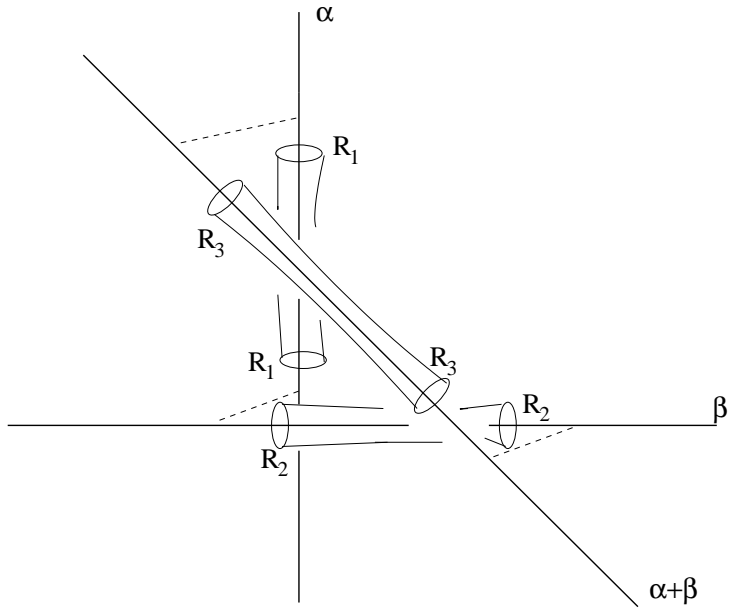}
\end{center}
\caption{The only nondegenerate instantons contributing to the
geometry depicted here are annuli stretching along the degeneracy
locus.}\label{annuli}
\end{figure}

The problem now is to determine how many configurations like this one
contribute to the full amplitude. It turns out that the only contributions
come from open strings stretching along the degeneracy locus. This was
found by Diaconescu, Florea and Grassi \cite{dfg2}
using localization arguments, and
derived in \cite{amv} by exploiting invariance under deformation of complex
structures. This result simplifies the problem enormously, and gives a
precise description of all the nondegenerate instantons contributing in
this geometry: they are annuli stretching along the fixed lines of the
${\bf T}^2$ action,
together with their multicoverings. This is illustrated in \figref{annuli}.
The action describing the dynamics of topological D-branes is then:
\be
\label{totalaction}
S=\sum_{i=1}^3 S_{\rm CS}(A_i) + \sum_{n=1}^{\infty}\; {1 \over n} \Bigl(
{\rm e}^{-nr_1}
\;{\rm Tr} U_1^{n}\; {\rm Tr}U_2^{-n}+ {\rm e}^{-nr_2}
\;{\rm Tr} U_2^{n}\; {\rm Tr}U_3^{-n}+ {\rm e}^{-nr_3}
\;{\rm Tr} U_3^{n}\; {\rm Tr}U_1^{-n} \Bigr),
\ee
where the $A_i$ are $U(N_i)$ gauge connections on each of the
${\bf S}^3$'s, $i=1,2,3$, and $U_i$ are the corresponding holonomies
around loops. There is a very convenient way to write the free energy of the
theory with the above action. First notice that, by following the same
steps that led to (\ref{ovrep}), one can write the operator
(\ref{massiveov})
as
\be
{\cal O}(U_1,U_2;r)=\sum_R {\rm Tr}_R
U_1 {\rm e}^{-\ell r}{\rm Tr}_R U^{-1}_2,
\ee
where $\ell$ denotes the number of boxes of the representation $R$. In the
situation depicted in \figref{annuli}, we see that there are two annuli
ending on each three-sphere. The boundaries of these annuli give knots, so
we have a two-component link in each ${\bf S}^3$. The holonomies around the
components of these links will be in different representations of $U(N)$,
as indicated in \figref{annuli}. Therefore, the free energy will be
given by:
\be
\label{freethree}
F =\sum_{i=1}^3 F_{\rm CS}(N_i, g_s) + \log \biggr\{ \sum_{R_1,R_2, R_3} {\rm e}^{-\sum_{i=1}^3 \ell_i r_i }
W_{R_1, R_2} ({\cal L}_1)  W_{R_2, R_3} ({\cal L}_2)
W_{R_3, R_1} ({\cal L}_3) \biggl\},
\ee
where $\ell_i$ is the number of boxes in the representation $R_i$, and
$F_{\rm CS}(N_i, g_s)$ denotes the free energy of Chern-Simons theory
with
gauge group $U(N_i)$. These correspond to the degenerate
instantons that come from each of the three-spheres.

\begin{figure}
\leavevmode
\begin{center}
\epsfysize=3cm
\epsfbox{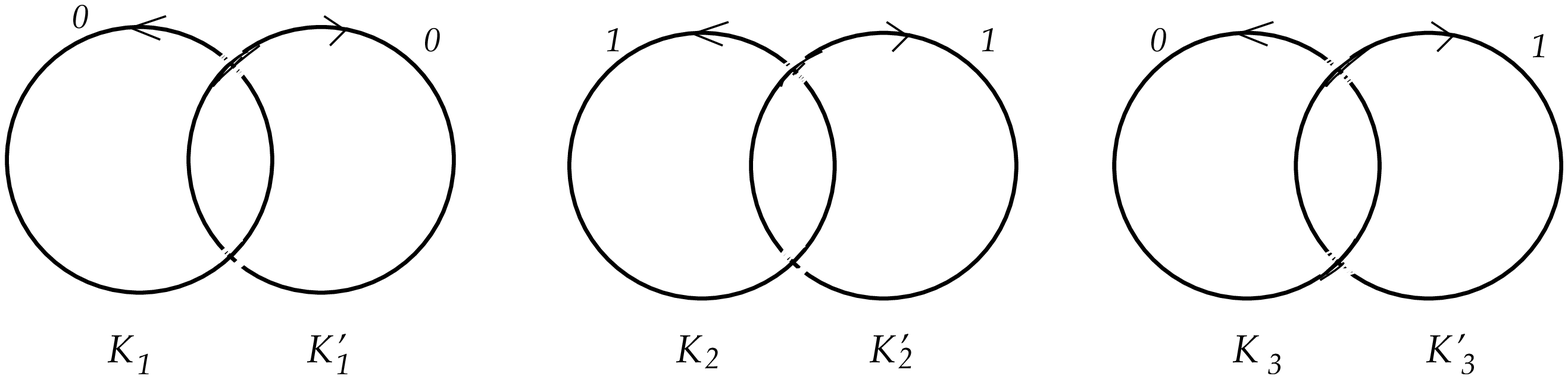}
\end{center}
\caption{The figure shows the Hopf links ${\cal L}_i$, $i=1,2,3$. The numbers
indicate the framing of each knot.}\label{hopfs}
\end{figure}

Of course, in order to compute (\ref{freethree})
we need some extra information: we have to know
what are, topologically, the links ${\cal L}_i$, and also
if there is some framing induced by the
geometry. It turns out that these questions can be easily answered by
looking at the geometry of the degeneracy locus. The key point is to note
that in this geometry the three-spheres represented by dashed lines between two
degeneracy loci have natural Heegard splittings into
two tori, and the gluing instructions are determined by the ${\rm Sl}(2,
{\bf Z})$ transformation
that maps the degenerating cycle at the end of the corresponding
three-sphere, to the degenerating cycle at the other end
\cite{amv}. For example, the three-sphere
between the $\alpha$ and the $\beta$
degenerating loci in \figref{annuli} comes from gluing two tori with an
$S^{-1}$ transformation, which maps the $\alpha$ cycle into the $\beta$
cycle. Following this procedure (see \cite{amv} for details) one finds
that the ${\cal L}_i$ are all Hopf links (see \figref{nudos}), and that some
of the components do actually have nontrivial framing.
If we denote the components of
${\cal L}_i$ by ${\cal K}_i$ and ${\cal
K}_i'$, $i=1,2,3$, the framings turn out to be the following:
${\cal K}_1$, ${\cal K}'_1$
and ${\cal K}_3$ have framing zero, while the remaining knots have framing
$p=1$. This means that ${\cal L}_1$ is in the canonical framing, in ${\cal
L}_2$ both components are framed, while in ${\cal L}_3$ only one of the
components, ${\cal K}'_3$, is framed. This is depicted in \figref{hopfs}.

What happens now if we go through the geometric transition of
\figref{threetrans}? As in the case originally studied by Gopakumar and
Vafa, the string coupling constant gives the Chern-Simons ``effective''
coupling constant $g_s=2\pi/(k_i + N_i)$ (which is the same for the three
theories, see (\ref{sameness})), while the 't Hooft parameters $t_i=g_s N_i$ correspond to the
sizes of the three outward legs of the toric diagram on the right side of
\figref{threetrans}. The free energy (\ref{freethree}) is, due to the
large $N$ transition, the free energy of topological {\it closed} strings
propagating in that toric geometry. In order to recover just local $\IP^2$, we have
to take the 't Hooft parameters to infinity, and ``tune'' the sizes of the
annuli at the same time.
It turns out that one has to perform a double scaling limit,
taking both $t_i$ and $r_i$ to infinity in such a way that
\be
\label{renorm}
r=r_1 -{ {t_1 + t_3} \over 2}=r_2 -{ {t_1 + t_2} \over 2}=
r_3 -{ {t_2 + t_3} \over 2}
\ee
remains finite. Then, $r$ can be identified with the complexified K\"ahler
parameter of local $\IP^2$. We refer again to \cite{amv} for details.
The free energy has in this limit the structure:
\be
F=\log \biggr\{ 1 + \sum_{\ell=1}^{\infty} a_{\ell} (q) {\rm e}^{-\ell
r}\biggr\}= \sum_{\ell=1}^\infty a_{\ell}^{(c)}(q) {\rm e}^{-\ell r}
\ee
where $q={\rm e}^{ig_s}$. The coefficients $a_{\ell} (q)$,
$a_{\ell}^{(c)}(q)$  can be easily
obtained in terms of
the invariants of the Hopf link in arbitrary representations. One finds,
for example \cite{amv},
\ben
a_1(q) &=& - { 3 \over (q^{-{1\over2}} -q^{1\over2})^2}, \nonumber\\
a_2^{(c)}(q)&=&
{ 6 \over (q^{-{1\over2}} -q^{1\over2})^2} + {1 \over 2}a_1 (q^2).
\een
If we compare to (\ref{gvseries}) and
take into account the effects of multicovering, we find the
following values for the Gopakumar-Vafa invariants of ${\cal O}(-3)
\rightarrow \IP^2$:
\ben
n_1^0=3, \,\,\,\,\,\,\,\,\,\,\,\,\,\,\,\,\,\,\,  n_1^g =0 \,\,\, {\rm
for}\, g>0,\nonumber\\
 n_2^0 = -6, \,\,\,\,\,\,\,\,\,\,\,\,\,\,\,\,\,\,\,  n_2^g =0 \,\,\, {\rm
for}\, g>0,
\een
in agreement with the results listed in Table \ref{Pgv}. In fact, one can
go much further with this method and compute the Gopakumar-Vafa invariants
to high degree. The advantage of this procedure is that, in contrast to
both the A and the B model computations, one gets the answer for {\it all
genera}, see \cite{amv} for a complete listing of the invariants up to
degree 12.

Although we have focused here on local $\IP^2$, one can analyze in a
similar
way other toric geometries, including local $\IP^1 \times \IP^1$ and other
local del Pezzo surfaces (see also \cite{dfg2, iqbal}). In fact, one can in
principle recover all local toric geometries in this way. We then see that
large $N$ transitions produce gauge theory duals of topological strings
propagating on various toric backgrounds. The gauge theory dual is given in
general by a product of Chern-Simons theories together
with complex scalars in
bifundamental representations, and moreover the gauge theory data are
nicely encoded in the toric diagram. Other aspects of these dualities for
toric
manifolds can be found in \cite{amv,dfg2,iqbal}.

\sectiono{Conclusions}

The remarkable connections between enumerative geometry and knot
invariants that have been reviewed in this paper certainly deserve further
investigation. Some directions for further research are the following:

1) The correspondence between knot invariants and open Gromov-Witten
   invariants has been tested only for the unknot. It would be very
interesting to test nontrivial knots and improve our understanding of the
map relating knots and links in ${\bf S}^3$ to Lagrangian submanifolds in
   the resolved conifold. This will certainly open new perspectives in
   the study of Chern-Simons knot invariants.

2) Another direction to explore is the correspondence between coupled
Chern-Simons systems and closed string invariants that we explained in
section 6. Extensions to more general local toric geometries, and even
to compact
geometries, would give a fascinating new point of view on the enumerative
geometry of Calabi-Yau threefolds.

3) The
``unreasonable effectiveness of physics in solving mathematical problems''
\cite{vafamath}
has given again surprising results connecting two seemingly unrelated
areas of geometry, and we need a deeper mathematical understanding of
these connections. For example, the results of section 6 may be understood in
terms of the localization techniques introduced in \cite{kont}, as
suggested in \cite{amv}.

\section*{Acknowledgments}
I would like to thank Mina Aganagic, Jose Labastida and
Cumrun Vafa for enjoyable collaborations on the topics
discussed in this review, and for sharing their insights with me.
I would also like to thank Andrew Neitzke for a careful reading of the
manuscript.
This work has been supported by
NSF-PHY/98-02709.

\end{document}